\documentclass[aoas, preprint]{imsartarxiv}

\RequirePackage[OT1]{fontenc}
\usepackage{subcaption} 
\usepackage[dvipsnames]{xcolor}
\usepackage{amsthm, amsmath, amsfonts, enumerate, mathtools, natbib}
\RequirePackage[colorlinks, citecolor=blue, urlcolor=blue]{hyperref}
\usepackage{verbatim} 
\usepackage{imsartarxiv}

\usepackage{chngcntr} 

\usepackage{tikz}
\usetikzlibrary{calc, fit, positioning, shapes,snakes}

\colorlet{highlight}{Black}
\colorlet{highlight2}{Black}

\usepackage{cleveref}
\crefformat{equation}{(#2#1#3)} 
\crefrangeformat{equation}{(#3#1#4) to~(#5#2#6)} 
\crefformat{enumi}{step~#2#1#3} 
\crefrangeformat{enumi}{steps~#3#1#4~--~#5#2#6}
\crefformat{figure}{Figure~#2#1#3}

\startlocaldefs
\numberwithin{equation}{section}
\theoremstyle{plain}

\long\def\@makecaption#1#2{%
  \vskip\abovecaptionskip
  \footnotesize
  \sbox\@tempboxa{\itshape\textsc{#1}. #2}%
  \ifdim \wd\@tempboxa >\hsize
    \itshape\textsc{#1}. #2\par
  \else
    \global \@minipagefalse
    \hb@xt@\hsize{\hfil\box\@tempboxa\hfil}%
  \fi
  \vskip\belowcaptionskip}

\long\def\@makecaption#1#2{%
  \vskip\abovecaptionskip
  \figurecaption@size
  \sbox\@tempboxa{{\figurename@size #1}\figurename@skip #2}%
  \ifdim \wd\@tempboxa >\hsize
    {\figurename@size #1}\figurename@skip #2\par
  \else
    \global \@minipagefalse
    \hb@xt@\hsize{\hfil\box\@tempboxa\hfil}%
  \fi
  \vskip\belowcaptionskip}

\setattribute{figurecaption}{size}{\footnotesize\itshape}
\setattribute{figurename}   {size}{\scshape}
\setattribute{figurename}   {skip}{.~~}

\def\@floatboxreset{%
        \reset@font
        \@setminipage
        \singlespacing
        \footnotesize
        \centering
}

\endlocaldefs

\newcommand{\beginsupplement}{%
        \setcounter{table}{0}
        \renewcommand{\thetable}{S\arabic{table}}%
        \setcounter{figure}{0}
        \renewcommand{\thefigure}{S\arabic{figure}}%
         \setcounter{section}{0}
        \renewcommand{\thesection}{S\arabic{section}}%
}

\newcommand{ \Prob  } {\mathbb{P} }
\newcommand{ \Exp  } {\mathbb{E} }
\newcommand{ \iid } { \stackrel{i.i.d.}{\sim} }
\newcommand{ \indep } { \stackrel{indep.}{\sim} }

\allowdisplaybreaks

\begin{document}

\begin{frontmatter}
\title{A Unified Statistical Framework for Single Cell and Bulk RNA Sequencing Data}
 \runtitle{A Unified Statistical Framework}
 
\begin{aug}
\author{\fnms{Lingxue} \snm{Zhu}\thanksref{t1}\ead[label=e1]{lzhu@cmu.edu}},
\author{\fnms{Jing} \snm{Lei}\thanksref{t1}\ead[label=e2]{jinglei@andrew.cmu.edu}}
\author{\fnms{Bernie} \snm{Devlin}\thanksref{t2}\ead[label=e3]{devlinbj@upmc.edu}}
\and
\author{\fnms{Kathryn} \snm{Roeder}\corref{}\thanksref{t1}\ead[label=e4]{roeder@andrew.cmu.edu}}

\affiliation{Carnegie Mellon University\thanksmark{t1}
and
University of Pittsburgh\thanksmark{t2}
}

\address{Department of Statistics\\
Carnegie Mellon University\\
5000 Forbes Avenue\\
Pittsburgh, Pennsylvania 15213\\
USA\\
\printead{e1}\\
\phantom{E-mail:\ }\printead*{e2}\\
\phantom{E-mail:\ }\printead*{e4}}

\address{Department of Psychiatry and Human Genetics\\
University of Pittsburgh School of Medicine\\
3811 O'Hara Street\\
Pittsburgh, Pennsylvania 15213\\
USA\\
\printead{e3}}

\runauthor{L. Zhu, J. Lei 
, B. Devlin 
and K. Roeder}
\end{aug}

\begin{abstract}

Recent advances in technology have enabled the measurement of RNA levels for individual cells. Compared to traditional tissue-level bulk RNA-seq data, single cell sequencing yields valuable insights about gene expression profiles for different cell types, which is potentially critical for understanding many complex human diseases. However, developing quantitative tools for such data remains challenging because of high levels of technical noise, especially the ``dropout" events.  A ``dropout" happens when the RNA for a gene fails to be amplified prior to sequencing, producing a ``false'' zero in the observed data. In this paper, we propose a Unified RNA-Sequencing Model (URSM) for both single cell and bulk RNA-seq data, formulated as a hierarchical model. URSM borrows the strength from both data sources and carefully models the  dropouts in single cell data, leading to a more accurate estimation of cell type specific gene expression profile. In addition, URSM naturally provides inference on the dropout entries in single cell data that need to be imputed for downstream analyses, as well as the mixing proportions of different cell types in bulk samples. We adopt an empirical Bayes approach, where parameters are estimated using the EM algorithm and approximate inference is obtained by Gibbs sampling. Simulation results illustrate that URSM outperforms existing approaches both in correcting for dropouts in single cell data, as well as in deconvolving bulk samples. We also  demonstrate an application to gene expression data on fetal brains, where our model successfully imputes the dropout genes and reveals cell type specific expression patterns.

\end{abstract}


\end{frontmatter}


\section{Introduction}
\label{sec:introduction}

A biological organism is made up of individual cells, which work in concert in tissues to constitute functioning organs. Biologists have long thought that the key to understanding most human diseases lies in understanding the normal and abnormal function of cells.  Yet, until very recently, our view of what molecules are expressed and where and when was limited to the level of tissues. Indeed RNA sequencing (RNA-seq) was introduced as a critical tool to answer these questions, but the RNA itself was collected from tissues. This bulk RNA-seq data provides reliable measurements of gene expression levels throughout the genome for bulk samples. With sufficient sequencing depth, even weakly expressed transcripts can be accurately captured by RNA-seq data.  This technology has led to breakthroughs in various fields.  For example, \cite{fromer2016gene} use bulk data, obtained from prefrontal cortex of post-mortem subjects, to gain insight into how genetic risk variation for schizophrenia affects gene expression and likely generates risk for this severe psychiatric disorder.  

Still bulk RNA-seq data inevitably ignores the heterogeneity of individual cells because the measurements are summed over the population of cells in the tissue. Yet it is reasonable to predict that diseases like schizophrenia do not arise from malfunctioning brain tissue, per se, but rather certain malfunctioning cells within that tissue. A leading hypothesis is that schizophrenia arises from synaptic dysfunction, and synapses are fundamental to neurons, so should neurons alone be targeted for analyses into schizophrenia? Actually, brain tissue is composed of a remarkably heterogeneous set of cell types, which have vastly different functions and expression profiles. While many are different types of neurons, many others support and alter the function of those neurons and their synapses. Thus, the different gene expression profiles for distinct cell types can have profound functional consequences. These likely are critical for the development of tissues and human diseases, and will be especially important as we aspire to fix such complex diseases as schizophrenia.

It is also of interest to link gene expression with genetic variation, particularly damaging variants associated with risk of disease.  Until recently researchers have assumed that most cells express both copies of a gene equally; however, new findings suggest an even more complex situation motivating single cell measurements.  Apparently some neurons preferentially express the copy of a gene inherited from one parent over the other and this can shape how mutated genes are expressed at the cellular level \citep{huang2017diverse}.

One approach to characterize cell type specific gene expression profiles is to perform deconvolution on bulk RNA-seq data. 
Consider an observed gene expression matrix $X \in \mathbb{R}^{N \times M}$ for $N$ genes in $M$ bulk samples, each containing $K$ different cell types. The goal of deconvolution is to find two non-negative matrices $\widetilde{A} \in \mathbb{R}^{N \times K}$ and $W \in \mathbb{R}^{K \times M}$, such that 
\begin{equation}
 X \approx \widetilde{A} W\,,
 \label{eq:factorization} 
\end{equation}
where each column of $W$ represents the mixing proportion of $K$ cell types in each bulk sample, and each column of $\widetilde{A}$ represents the average gene expression levels in each type of cells. 
\textcolor{highlight}{
If the ``signature" matrix $\widetilde{A}$ is available for a set of ``barcode genes" in each cell type, deconvolution reduces to a regression-type problem that aims at estimating $W$. Several algorithms have been proposed under this setting, including Cibersort \citep{newman2015robust} and csSAM \citep{shen-orr2010cell}.} 
However, without knowing the signature matrix, deconvolution is highly nontrivial, and this line of methods includes the Deconf algorithm \citep{repsilber2010biomarker}, semi-supervised Nonnegative Matrix Factorization algorithm (ssNMF) \citep{gaujoux2012semi}, and Digital Sorting Algorithm (DSA) \citep{zhong2013digital}. 

A fundamental challenge of the NMF-based methods is the non-uniqueness of the factorization \citep{donoho2003does}. Therefore, to obtain a biologically meaningful result, both ssNMF \citep{gaujoux2012semi} and DSA \citep{zhong2013digital} use a set of ``marker genes'' to guide the factorization. A marker gene is a gene that only expresses in one cell type. In other words, there are several rows of $\widetilde{A}$ that are priorly known to be non-zero at only one column. This is equivalent to the separability assumption introduced by \cite{donoho2003does} for the uniqueness of NMF. Unfortunately, marker genes are rarely known in practice. In fact, extracting high-quality marker genes is a challenging step, which is often approached by analyzing purified cells \citep{abbas2009deconvolution}.

On the other hand, single cell RNA sequencing provides gene expression measurements in individual cells, yielding a high-resolution view of cellular states that are uncharacterized in bulk data. Recent advances in high-throughput technologies have made it possible to profile hundreds and thousands of cells \citep{kolodziejczyk2015technology, fan2015combinatorial}.  With several extra pre-processing steps including reverse transcription and amplification, the single cell mRNA library goes through similar sequencing procedures as the bulk samples, and the gene expression levels are measured by the number of mapped reads.  With single cell RNA-seq data, one can investigate distinct subpopulations of cells, gain better understanding of the developmental features of different cell types \citep{grun2015single}, identify cellular differences between healthy and diseased tissues \citep{kharchenko2014bayesian}, and infer gene-regulatory interactions \citep{padovan2013using}.

The challenges of modeling single cell RNA-seq data come from high cell-to-cell variation, as well as high levels of technical noise during sequencing due to the low amounts of starting mRNAs in individual cells. One important bias comes from the so-called ``dropout'' events. A dropout happens when a transcript is not detected due to failure of amplification prior to sequencing, leading to a ``false'' zero in the observed data \citep{kolodziejczyk2015technology}. Given the excessive amount of zero observations in single cell RNA-seq data, it is critical to distinguish between (i) the dropout genes where transcripts are missed in sequencing; and (ii) the ``structural zeros'' where the genes are truly un-expressed. Modeling the dropout events is especially challenging because of their complicated dependency on gene expression levels and cell characteristics. Specifically, dropouts are more likely to occur in genes expressed at low levels, and certain cells may have systematically higher dropout probabilities than others. 
 In addition to dropout events, other challenges in modeling single cell data include the over-dispersion due to both cellular and technical variation, as well as high magnitude outliers due to bursts and fluctuations of gene expression levels.
 \textcolor{highlight2}{We refer the readers to \cite{haque2017a-practical} for a more comprehensive review.}

Despite the success of many early single-cell studies, statistical tools that account for the technical noise in single cell RNA-seq data, especially the dropout events, are limited.  There have been efforts to analyze single cell data for various purposes. Many methods propose to quantify and account for technical noise using spike-ins \citep{brennecke2013accounting, vallejos2015basics:, vallejos2016beyond}. However, spike-ins are usually unavailable in single cell data due to its expenses in practice. For differential expression analysis, SCDE \citep{kharchenko2014bayesian} is based on a Bayesian hypothesis testing procedure using a three-component mixture model to capture technical noise; subsequently, MAST \citep{finak2015mast:} uses a hurdle model that can adjust for various covariates; more recently, \cite{vu2016beta-poisson} construct a beta-poisson mixture model, integrated within a generalized linear model framework. Various relevant problems have also been studied, including
 inferring the spatial localization of single cells in complex tissues \citep{satija2015spatial}, dimension reduction using Zero-Inflated Factor Analysis (ZIFA) \citep{pierson2015zifa}, and clustering unlabeled single cells while accounting for technical variation \citep{prabhakaran2016dirichlet}.  All of these aforementioned methods have been successfully applied to different single cell data sets.
However, analytical methods that aim at the fundamental problem of imputing dropout genes and estimating the cell-type-specific gene expression profiles remain underdeveloped. 

In this paper, we propose to jointly analyze single cell and bulk RNA-seq data using the Unified RNA-Sequencing Model (URSM), which simultaneously corrects for the dropout events in single cell data and performs deconvolution in bulk data. 
\textcolor{highlight}{
We point out that URSM only requires consistent cell types between both data sources, preferably measured on the same tissue from subjects with similar ages. It does not require the single cell and bulk data being measured on the same subjects, nor does it assume the same proportions of cell types in both data sets. Given a single cell data set, usually there are existing bulk data measured on the same tissue that can be modeled jointly using URSM. For example, BrainSpan provides extensive gene expression data on adult and developing human brains \citep{sunkin2013allen}, and GTex establishes a human RNA-seq gene expression database across 43 tissues \citep{gtex-consortium2013the-genotype-tissue}.
}

By integrating single cell and bulk RNA-seq data, URSM borrows the strength from both data sources, and is able to (i) obtain reliable estimation of cell type specific gene expression profiles; (ii) infer the dropout entries in single cell data; and (iii) infer the mixing proportions of different cell types in bulk samples. Our framework explicitly models the dropout events in single cell data, and captures the relationship between dropout probability and expected gene expression levels. By involving high-quality bulk data, URSM achieves more accurate estimation of cellular expression profiles than using only single cell data. By incorporating the single cell data, URSM provides, for the first time, deconvolution of the bulk samples without going through the error-prone procedure of estimating marker genes. To the best of our knowledge, this is the first model that jointly analyzes these two types of RNA-seq data.  We will illustrate in simulation (\Cref{sec:simulation}) and real-world data (\Cref{sec:data}) that URSM successfully corrects for the dropouts in single cell data, and provides reliable deconvolution for bulk samples.


\section{A Unified Statistical Model}
\label{sec:model}

Suppose RNA-sequencing is conducted on $N$ genes and $K$ types of cells are of interest. Then bulk and single cell RNA-seq data can be linked together by a common profile matrix $A \in \mathbb{R}^{N \times K}$, where the $k$-th column $A_{\cdot k}$ represents the expected {\it relative} expression levels of $N$ genes in the $k$-th type of cells, such that each column sums to one. Note that by considering the {\it relative} expression levels, the profile matrix $A$ does not depend on sequencing depths, and thus remains the same in both data sources. The two data sources provide two different views on the profile matrix $A$. In single cell data, the observations are independent realizations of different columns of $A$ with extra noise due to dropout events. In bulk data, the expected relative expression levels for a mixture sample are weighted sums of columns of $A$, where the weights correspond to mixing proportions of different cell types. 
Here, we propose URSM to analyze the bulk and single cell RNA-seq data together, which borrows the strength from both data sets and achieves more accurate estimation on the profile matrix. This further enhances the performance of deconvolving bulk samples, as well as inferring and imputing the dropout genes in single cells.  

\begin{figure}

\begin{center}
\begin{tikzpicture}[node distance=8mm]
\tikzstyle{main}=[circle, minimum size = 6mm, thick, draw =black!80, node distance = 9mm]
\tikzstyle{parameter}=[diamond, minimum size = 8mm, thick, draw =black!80, node distance = 16mm]

\tikzstyle{connect}=[-latex, thick]
\tikzstyle{box}=[rectangle, draw=black!100]

   \node[main, fill = black!10] (Y) [label=center:$Y_{\cdot l}$] { };
    \node[main] (S) [left=of Y, label=center:$S_{il}$] {};  
   \node[main, double] (Pi) [left=of S,label=center:$\pi_{il}$] { };
    \node[main] (kappa) [left=of Pi, yshift=4mm, label=center:${\kappa_l}$] { };
    \node[main] (tau) [left=of Pi, yshift=-4mm, label=center:${\tau_l}$] { };
    \node (kappaPara) [left=of kappa] {$\mu_\kappa, \sigma_\kappa^2$};
    \node (tauPara) [left=of tau] {$\mu_\tau, \sigma_\tau^2$};
    
    \node (A) [right=of Y, yshift=10mm] {$A$};
    
    \node[main, fill = black!10] (X) [right=of A, yshift=-10mm, label=center:$X_{\cdot j}$] { };
    \node[main] (W) [right=of X, label=center:$W_{\cdot j}$] {};  
    \node (alpha) [right=of W] {$\alpha$};
        
  \path 
  	(S) edge [connect] node [below]  { \textcolor{blue}{\it\scriptsize Mult} } (Y)
        (Pi) edge [connect] node [below] { \textcolor{blue}{\it\scriptsize Bern} }  (S)
        (A) edge [connect] node [sloped, yshift=-2mm, xshift=1mm]  { \textcolor{blue}{\it\scriptsize Mult} } (Y)
        (A) edge [connect] node [sloped, yshift=-2mm, xshift=-2.2mm]  { \textcolor{blue}{\it\scriptsize Mult} } (X)
        (W) edge [connect] node [above]  { \textcolor{blue}{\it\scriptsize Mult} }  (X)
        (alpha) edge [connect] node [above, xshift=2mm]  { \textcolor{blue}{\it\scriptsize Dir} }  (W)
        (A) edge [connect] node [sloped, anchor=center, above, xshift=-12mm, yshift=-0.5mm] { \textcolor{blue}{\it\scriptsize logistic} } (Pi)
        (kappa) edge [connect] node [below, yshift=0.5mm, xshift=-1.5mm] { \textcolor{blue}{\it\scriptsize logistic} } (Pi)
        (tau) edge [connect] (Pi)
        (kappaPara) edge [connect] node [above, xshift=-2mm]  { \textcolor{blue}{\it\scriptsize N} }  (kappa)
        (tauPara) edge [connect] node [above, xshift=-2mm]  { \textcolor{blue}{\it\scriptsize N} }  (tau)
        ;
  
  \node[rectangle, inner sep=0mm, fit= (S), label=below right:$N$, yshift=0mm, xshift=-1mm] {};
  \node[rectangle, inner sep=4mm,draw=black!100, fit= (S) (Pi)] {};
 
  \node[rectangle, inner sep=0mm, fit= (Y), label=below right:$L$, xshift=-2mm, yshift=-1.5mm] {};
  \node[rectangle, inner sep=2mm, draw=black!100, fit = (Y) (kappa) (tau)] {};
  
   \node[rectangle, inner sep=0mm, fit= (W), label=below right:M, xshift=-1mm, yshift=1mm] {};
  \node[rectangle, inner sep=4mm, draw=black!100, fit = (X) (W)] {};

\end{tikzpicture}
\end{center}
\caption{Plate model of URSM, with both single cell data (on the left) and bulk samples (on the right). The two greyed nodes $X$  and $Y$ represent observed gene expression levels. Node $S$ is a binary variable representing dropout status in single cells, and node $W$ represents the mixing proportions in bulk samples. The node $\pi$ representing observation probability is double-circled because it is deterministic, and all model parameters are shown without circles, including the profile matrix $A$ that links the two data sources.}
\label{fig:plate}
\end{figure}
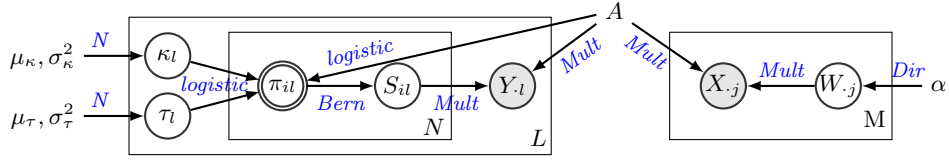

The plate model of URSM for generating single cell and bulk RNA-seq data is given in Figure~\ref{fig:plate}. Specifically, for single cell data, let $Y \in \mathbb{R}^{N \times L}$ represent the measured expression levels of $N$ genes in $L$ single cells, where the entries are RNA-seq counts. To model the dropout events, we introduce the binary observability variable $S \in \{0, 1\}^{N \times L}$,  where $S_{il}=0$ if gene $i$ in cell $l$ is dropped out, and $S_{il}=1$ if it is properly amplified. For each cell $l$, let $G_l \in \{1, \cdots, K\}$ denote its type, then the vector of gene expression $Y_{\cdot l} \in \mathbb{R}^N$ is assumed to follow a Multinomial distribution 
\textcolor{highlight}{
with probability vector $p_l$,} 
and the sequencing depth $R_l = \sum_{i=1}^N Y_{il}$ is the number of trials. 
\textcolor{highlight}{
Without dropout events, $p_l$ would be the corresponding column of the profile matrix, $A_{\cdot  \,G_l}$, which is the true relative expression levels for cell type $G_l$. With the existence of dropouts, $p_l$ becomes the element-wise product of $A_{\cdot \, G_l}$ and $S_{\cdot l}$, which is then normalized to sum to one. 
To capture the dependency between dropout probabilities and gene expression levels, the observation probability $\pi_{il}=\mathbb{P}(S_{il}=1)$ is modeled as a logistic function of $A_{i, G_l}$,
\begin{equation}
\pi_{il} = \textrm{logistic}\left( \kappa_l + \tau_l  A_{i, G_l}  \right)\,,
\end{equation}
so that lowly expressed genes have high probabilities of being dropped out, where the coefficients $(\kappa_l, \tau_l)$ are cell-dependent that capture the cellular heterogeneity. 
}
Under this model, the set of dropout entries and structural zeros are defined as
\begin{equation}
\begin{split}
&\textrm{dropouts} = \{(i, l): S_{il}=0\}\,, \\
& \textrm{structural zeros} = \{(i, l): S_{il}=1, Y_{il}=0\}\,.
\end{split}
\label{eq:drop-struct}
\end{equation}

For bulk data, let $X \in \mathbb{R}^{N \times M}$ represent the RNA-seq counts of $N$ genes in $M$ bulk samples. For the $j$-th bulk sample, let $W_{\cdot j} \in \mathbb{R}^K$ denote the mixing proportions of $K$ cell types in the sample, satisfying $\sum_{k=1}^K W_{kj} = 1$. Then the gene expression vector $X_{\cdot j} \in \mathbb{R}^N$ is assumed to also follow a Multinomial distribution, where the probability vector is the weighted sum of $K$ columns of $A$ with the weights being $W_{\cdot j}$, 
\textcolor{highlight}{
and the number of trials is  the sequencing depth for sample $j$, defined as $R_j = \sum_{i=1}^N X_{ij}$. 
}

For the hierarchical model setting, we assign the conjugate Dirichlet prior for the mixing proportions $W_{\cdot j}$, and Gaussian priors for the cell-dependent dropout parameters $(\kappa_l, \tau_l)$. 
Here, we adopt an empirical Bayes approach, where the parameters are estimated by maximum-likelihood-estimations (MLE) using the expectation-maximization (EM) algorithm.  
Using this framework, our goal is threefold: (i) learn the profile matrix $A$ as part of the model parameters, which characterizes the cellular gene expression profiles; (ii) make posterior inference on the dropout status $S$ for single cell data, which can be used to identify dropout entries, and (iii) make posterior inference on the mixing proportions $W$ in bulk samples. Finally, the inferred dropout entries in single cell data can be imputed by their expected values using the estimated $A$ 
and sequencing depths $R_l$.


\paragraph{Full model specification.} 
\begin{itemize}
\item Bulk data
\begin{itemize}
\item $W_{\cdot j} \iid \textrm{Dirichlet} (\alpha)$ for $j=1, \cdots, M$, where $\alpha \in \mathbb{R}^K$, $\alpha \geq 0$.

\item $X_{\cdot j}  \,|\,  W_{\cdot j} \indep \textrm{Multinomial}(R_j, ~ A W_{\cdot j } )$ for $j=1, \cdots, M$, where $R_j = \sum_{i=1}^N X_{ij}$.
\end{itemize}

\item Single cell data
\begin{itemize}
\item $\kappa_l \iid N(\mu_{\kappa}, \sigma_{\kappa}^2)$, $\tau_l \iid N(\mu_{\tau}, \sigma_{\tau}^2)$ for $l=1, \cdots, L$.
\item $\pi_{il} = \textrm{logistic}\left( \kappa_l + \tau_l  A_{i, G_l}  \right)$, where $G_l \in \{1, \cdots, K\}$ is the type of the $l$-th cell.
\item $S_{il}  \,|\,  \kappa_l, \tau_l \indep \textrm{Bernoulli}(\pi_{il})$ for $i = 1, \cdots, N; ~ l=1, \cdots, L$.

\item $Y_{\cdot l}  \,|\,  S_{\cdot l} \indep \textrm{Multinomial}( R_l, ~ p_{l})$ for $l=1, \cdots, L$, where $R_l= \sum_{i=1}^N Y_{il}$, 
\[ p_l = (p_{il})_{i = 1, \cdots, N}\,, \textrm{ where } p_{il} = \frac{  A_{i, G_l}  S_{il} }  { \sum_{n=1}^N A_{n, G_l}  S_{nl} }\,. \]

\end{itemize}
\end{itemize}

\paragraph{Remark 1} We assume all entries in $A$ to be strictly positive. In principle, one can allow some entries $A_{ik}$ to be exactly zero, but this will lead to a degenerate multinomial distribution and complicate the likelihood function. In addition, making inference on $S_{il}$ when $A_{i, G_l}=0$ is an ill-defined problem. If $A_{ik}=0$, then we will have $X_{il}=0$ for all type-$k$ cells, but such structure rarely appears in real data. In practice, it is usually helpful to use some small positive numbers rather than exact zeros to capture the background signal in sequencing processes \citep{kharchenko2014bayesian}.

\paragraph{Remark 2} It is straightforward to use one part of URSM when only one data source is available. In \Cref{sec:simulation}, we will show the performance of the submodel for single cell data. It is also possible to use the submodel for bulk data when only bulk data are available, but extra information about marker genes needs to be incorporated in this scenario to avoid the non-identifiability issue, as explained in \Cref{sec:introduction}.


\section{Inference and Estimation: EM Algorithm}
\label{sec:EM}

This section presents an expectation-maximization (EM) algorithm \citep{dempster1977maximum} for fitting the maximum likelihood estimation (MLE) of the parameters  $\theta=(A, \alpha, \mu_\kappa, \sigma_\kappa^2, \mu_\tau, \sigma_\tau^2)$, as well as a Gibbs sampling algorithm for posterior inference on latent variables $H = (W, S, \kappa, \tau)$.  As illustrated in \Cref{sec:model}, the key values of scientific interests include (i) an estimate of the profile matrix $A$ that characterizes the cellular  gene expression profiles; (ii) $\Exp[S | Y, \theta]$, the inferred dropout probability at each entry in single cell data; and (iii) $\Exp[W | X, \theta]$, the inferred mixture proportion of bulk samples. 

The main difficulty of handling our model is the intractable posterior distributions due to non-conjugacy.  Therefore, approximate inference needs to be performed. One of the main methods for approximate inference in Bayesian modeling is Monte Carlo Markov Chain (MCMC) sampling \citep{alan-gelfand1990sampling-based}, where a Markov chain on latent variables is constructed, with stationary distribution being the true posterior.  After obtaining a long enough chain, the posterior can be approximated with empirical estimation. Gibbs sampling \citep{geman1984stochastic, casella1992explaining} is one of the most widely used forms of MCMC algorithms given its simplicity and efficiency.  On the other hand, variational methods form an alternative line for approximate inference, where the posterior is approximated analytically by a family of tractable distributions \citep{jordan1999introduction, wainwright2008graphical, blei2016variational}. While being computationally scalable in many large-scale problems, variational methods are inherently less accurate due to the  inevitable gap between the variational distributions and the true posterior distribution.

In this paper, we present a Gibbs sampling algorithm for approximate inference on latent variables using the data augmentation trick. This algorithm can also be used in the E-step of the EM procedure, leading to a Gibbs-EM (GEM) algorithm for obtaining MLEs of model parameters \citep{dupuy2016online}. The specific steps are outlined in \Cref{sec:gibbs} and \Cref{sec:m-step}, and more details can be found in the supplement. Finally, we point out that one can also proceed with variational inference, but due to space limitation, we do not pursue this approach in detail.

\subsection{E-step: Gibbs sampling}
\label{sec:gibbs}
The latent variables for bulk data and single cell data are conditionally independent given observed data $X, Y$ and parameters. Therefore, Gibbs sampling can be performed on the two data sources in  parallel. In this section, we describe the sampling procedure for the two parts separately.

\paragraph{Bulk data}
To obtain the posterior inference of $W$ (the mixing proportions) in bulk data, we re-write the model to be mixture of multinomials by introducing the augmented latent variables $Z$ and $d$ as follows:
\begin{align}
\begin{split}
&W_{\cdot j} \iid \textrm{Dirichlet}(\alpha)\,,~ j=1, \cdots, M\,, \\
&Z_{rj} \iid \textrm{Multinomial}(1, W_{\cdot j})\,, ~ r=1, \cdots, R_j \,, \\
& d_{rj} \indep \textrm{Multinomial}(1, A_{\cdot Z_{rj}})\,, ~ r=1, \cdots, R_j\,, \\
& X_{ij} = \sum_{r=1}^{R_j} I_{\{d_{rj}=i\}}\,, ~ i=1, \cdots, N, ~ j=1, \cdots, M\,.
\end{split}
\end{align}
Note that this model is closely related to the Latent Dirichlet Allocation (LDA) model \citep{blei2003latent} in topic modeling, if we view a gene as a word, a cell type as a topic, and a bulk sample as a document. Although  the Gibbs sampling algorithm has been developed for LDA in \cite{griffiths2004finding}, there are two difficulties that prevent us from directly applying this algorithm to our model. First, the LDA model assumes observations of $d_{rj}$, which are the actual words in an document, but in RNA-seq data, only the final counts $X_{ij}$ are observed. Second, the sequencing depths $R_j$'s are typically large in real data, so it will be extremely computationally demanding to keep track of $Z_{rj}$ and $d_{rj}$. Therefore, we propose a modified algorithm by defining another set of augmented latent variables
\begin{equation}
 \tilde{Z}_{ij, k} := \sum_{r: d_{rj}=i} I_{\{Z_{rj}=k\}} ~\textrm{and}~ \tilde{Z}_{ij} := (\tilde{Z}_{ij, k} ) \in \mathbb{R}^K\,, 
 \end{equation}
and it can be shown that
\begin{align}
\begin{split}
& W_{\cdot j} \,|\, W_{\cdot (-j)}, \tilde{Z}, X \sim \textrm{Dirichlet} \left( \alpha + \sum_{i=1}^N \tilde{Z}_{ij} \right)\,, \\
& \tilde{Z}_{ij} \,|\,  \tilde{Z}_{(-ij)}, W, X \sim \textrm{Multinomial} \left( X_{ij}, \frac{ A_{i \cdot} \odot W_{\cdot j} }{ \sum_{k=1}^K A_{ik} W_{kj}}  \right)\,,
\end{split}
\end{align}
where $\odot$ denotes element-wise multiplication, and the index $(-i)$ denotes everything else other than $i$. 

\paragraph{Single cell data}
As for posterior inference of $S, \kappa, \tau$ in single cell data, note that the first part of the model can be re-written as
\begin{align}
\begin{split}
 & (\kappa_l, \tau_l) \sim N( \mu, \Sigma ) ,
 ~\textrm{where } \mu=(\mu_{\kappa}, \mu_{\tau}), \Sigma=\textrm{Diag}(\sigma_{\kappa}^2, \sigma_{\tau}^2)\,, \\
& S_{il}  \,|\,  \kappa_l, \tau_l \sim \textrm{Bernoulli}(\textrm{logistic}\left( \psi_{il}  \right))\,,~\textrm{where } \psi_{il} = \kappa_l + \tau_l A_{i, G_l}\,,
 \label{eq:bayes-logit}
 \end{split}
\end{align}
which has the same form as a Bayesian logistic regression, with covariates being $(1, A_{i, G_l})$. Therefore, following the recent development of Gibbs sampling technique in this area \citep{polson2013bayesian}, we introduce a set of augmented latent variables $\omega$, and the conditional complete posteriors can be shown to be
\begin{align}
\begin{split}
& \omega_{il} \,|\, \omega_{(-il)}, S, Y, \kappa, \tau ~ \sim ~ \textrm{Polya-Gamma}(1, \psi_{il})\,, \\
& (\kappa_l, \tau_l) \,|\, \kappa_{(-l)}, \tau_{(-l)}, \omega, S, Y  ~ \sim ~  N(m_{\omega l}, V_{\omega l}^{-1})\,, \\
& S_{il} \,|\, S_{(-il)}, \omega, S, \kappa, \tau, Y ~ \sim ~ \textrm{Bernoulli}(b_{il})\,,
\end{split}
\end{align}
where
\[\begin{split}
&\psi_{il} = \kappa_l + \tau_l A_{i, G_l}\,,\\
&V_{\omega l} = \left(\begin{array}{cc}
\sum_{i=1}^N \omega_{il} + \sigma_{\kappa}^{-2} & \sum_{i=1}^N \omega_{il} A_{i, G_l} \\
\sum_{i=1}^N \omega_{il} A_{i, G_l} & \sum_{i=1}^N \omega_{il} A_{i, G_l}^2 + \sigma_{\tau}^{-2}
\end{array}\right)
\,, \\
& m_{\omega l} = V_{\omega l}^{-1} 
 \left(\begin{array}{c}
 \sum_{i=1}^N S_{il} - N/2 + \mu_\kappa / \sigma_\kappa^2 \\
 \sum_{i=1}^N S_{il} A_{i, G_l} - 1/2 + \mu_\tau / \sigma_\tau^2 
 \end{array}\right)
 \,,\\
&b_{il} = \begin{cases}
1, & \textrm{if } Y_{il} > 0 \\
\textrm{logit}\left( \psi_{il} + R_l \log \left( \frac{\sum_{n \neq i} A_{n, G_l}S_{nl}}{A_{i, G_l}  + \sum_{n \neq i} A_{n, G_l}S_{nl}} \right) \right), & \textrm{if } Y_{il}=0
\end{cases}
\,.
\end{split}\]

\subsection{M-step}
\label{sec:m-step}
In the M-step of GEM algorithm, the parameters are updated to maximize a lower bound on the expected complete log likelihood function, or the so-called Evidence Lower BOund (ELBO), where the posterior expectation $\Exp_Q$ is estimated using Gibbs samples obtained in the E-step. The optimal dropout parameters $(\mu_\kappa, \sigma_\kappa^2, \mu_\tau, \sigma_\tau^2)$ have the following closed forms: 
\begin{align}\begin{split}
\hat{\mu}_\kappa = \frac{1}{L} \sum_{l=1}^L \Exp_Q(\kappa_l)\,, ~~ \hat{\sigma}_\kappa^2 = \frac{1}{L} \sum_{l=1}^L \Exp_Q \left[( \kappa_l - \hat{\mu}_\kappa )^2 \right]\,, \\
\hat{\mu}_\tau = \frac{1}{L} \sum_{l=1}^L \Exp_Q(\tau_l)\,, ~~ \hat{\sigma}_\tau^2 = \frac{1}{L} \sum_{l=1}^L \Exp_Q \left[( \tau_l - \hat{\mu}_\tau )^2 \right]\,.
\end{split}\end{align}
For $A$ and $\alpha$, there are no closed form solutions, and we use the projected gradient ascent algorithm:
\begin{equation}
\begin{split}
& A_{\cdot k}^{new} \leftarrow \textrm{Proj} \left( A_{\cdot k}^{old} + t \cdot \nabla ELBO( A_{\cdot k}^{old} ) \right) \,, \\
& \alpha^{new} \leftarrow\textrm{Proj} \left(  \alpha^{old} + t \cdot \nabla ELBO(\alpha^{old}) \right) \,,
\label{eq:gd}
\end{split}
\end{equation}
where the step size $t$ is determined by backtracking line search, and the \textsl{Proj} function is the projection onto the feasible set:
\begin{equation}
A_{ik} \geq \epsilon_A, ~~ \sum_{i=1}^N A_{ik} = 1, ~~ \alpha_k \geq \epsilon_{\alpha}\,,
\end{equation}
where $\epsilon_A,  \epsilon_{\alpha} > 0$ are some small pre-determined constants.  The gradients are computed as
\begin{equation}
\begin{split}
 & \frac{ \partial ELBO}{ \partial A_{ik} } = \sum_{j=1}^M \frac{ \Exp_Q\left[ \widetilde{Z}_{ij,k} \right] }{A_{ik}} + 
 \sum_{l: G_l=k}\left[ \frac{  Y_{il} \Exp_Q(S_{il}) }{ A_{ik} } - \Exp_Q[ \omega_{il} \tau_l^2 ] A_{ik} - \right.  \\
& \qquad \qquad \qquad \left.  \frac{\Exp_Q(S_{il})  R_l}{ u_l} + \Exp_Q \left[ \left(S_{il} - \frac{1}{2} \right) \tau_l -\omega_{il} \tau_l \kappa_l \right]  \right] \,,   \\
 &  \frac{ \partial ELBO}{ \partial \alpha_k } = \sum_{j=1}^M \Exp_Q[ \log W_{kj} ] + M \left[ \Psi \left(\sum_{k=1}^K \alpha_k \right) - \Psi(\alpha_k) \right] \,,
 \end{split}
  \end{equation}
where $u_l = \sum_{i=1}^N A_{i, G_l} \Exp_Q(S_{il})$ and $\Psi(\cdot)$ is the digamma function. More detailed derivations can be found in the supplement.


\section{Simulation Results}
\label{sec:simulation}

In this section, we evaluate the performance of URSM in synthetic datasets. We focus on the accuracy of recovering the profile matrix $A$ and mixing proportions $W$, as well as the ability of distinguishing between dropout entries and structural zeros using the posterior inference of $S$. 

\subsection{Settings}
\label{sec:sim-setting}

\textcolor{highlight}{
Let $N$ be the number of genes. The sequencing depths for bulk samples are independently generated from Poisson($50 \, N$). To account for the fact that the sequencing depths of single cell data are usually much lower and highly variable, they are generated from Negative Binomial with mean $2 \, N$ and dispersion parameter $2$. 
}

\textcolor{highlight}{
The cell-type specific profile matrix $A$ is generated as follows:
(i) simulate all entries independently from log-normal with $\mu=0, \sigma=1$;
(ii) for each cell type $k$, let $N_m = 10 $ genes be marker genes, i.e., set $A_{il}=0$ for $l \neq k$;
(iii) for each cell type $k$, let $N_a = 10$ genes be anti-marker genes,  i.e., set $A_{ik}=0$;
(iv) let another set of $N_h = 30$ genes be house-keeping genes that have same expression levels in all cell types; 
(v) finally, properly normalize $A$ so that each column sums to 1. Specifically, in each column, we normalize the $N_h$ house-keeping genes such that they sum to $N_h/N$, and the remaining genes sum to $1 - N_h / N$.
}

\textcolor{highlight}{
Finally, the observation status $\{S_{il}\}_{il}$ for each gene $i$ in each single cell $l$ is simulated independently from Bernoulli$(\pi_{il})$. Recall that $S_{il}=0$ indicates a dropout, and the dropout probability is determined by 
\begin{equation}
1 - \pi_{il} = 1 - \textrm{logistic}(\kappa_l + \tau_l A_{i, G_l})\,,
\label{eq:dropout-curve}
 \end{equation}
where $G_l \in \{1, ..., K\}$ is the type of cell $l$. In the following sections, $\kappa_l$'s are independently generated from $\textrm{Normal}(-1, 0.5^2)$, and $\tau_l$'s are independently generated from $\textrm{Normal}(1.5 \, N, (0.15 \, N)^2)$. Note that by construction, the mean of each column of $A$,  $\bar{A}_{\cdot, k}$, is always $1/N$. Therefore, $\Exp[\kappa_l  + \tau_l \bar{A}_{\cdot, G_l}] = 0.5$ for each cell, which corresponds to an average dropout probability of $37.8\%$, and the maximal dropout probability is $73.1\%$ when $A_{ik}=0$.
}

\subsection{Estimation of profile matrix}
\label{sec:motivate}
\begin{figure}	
	\centering
	\begin{subfigure}[t]{0.24 \textwidth}
		\centering
		\caption{Dropout prob.}\label{fig:dropout-curve}	
		\includegraphics[width=\textwidth]{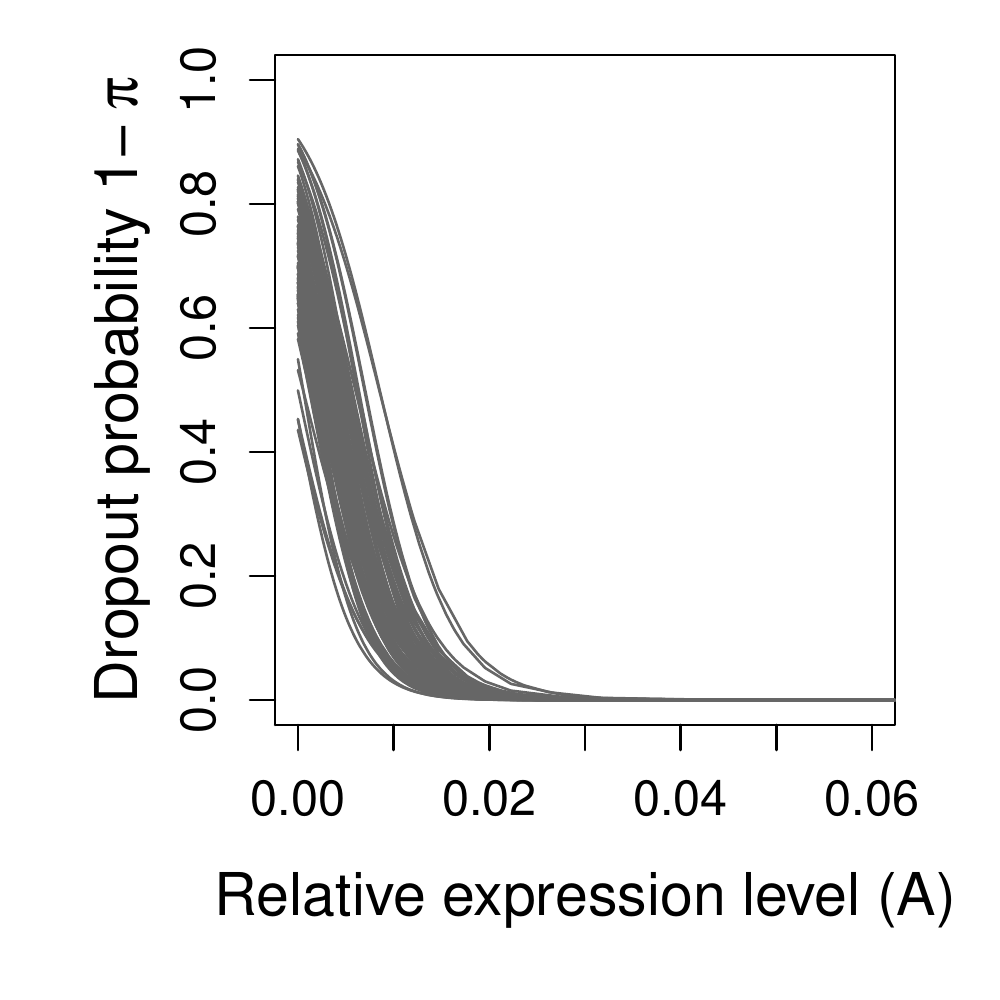}		
	\end{subfigure}
	\begin{subfigure}[t]{0.24 \textwidth}
		\centering
		\caption{Sample mean}\label{fig:sample-estA}
		\includegraphics[width=\textwidth]{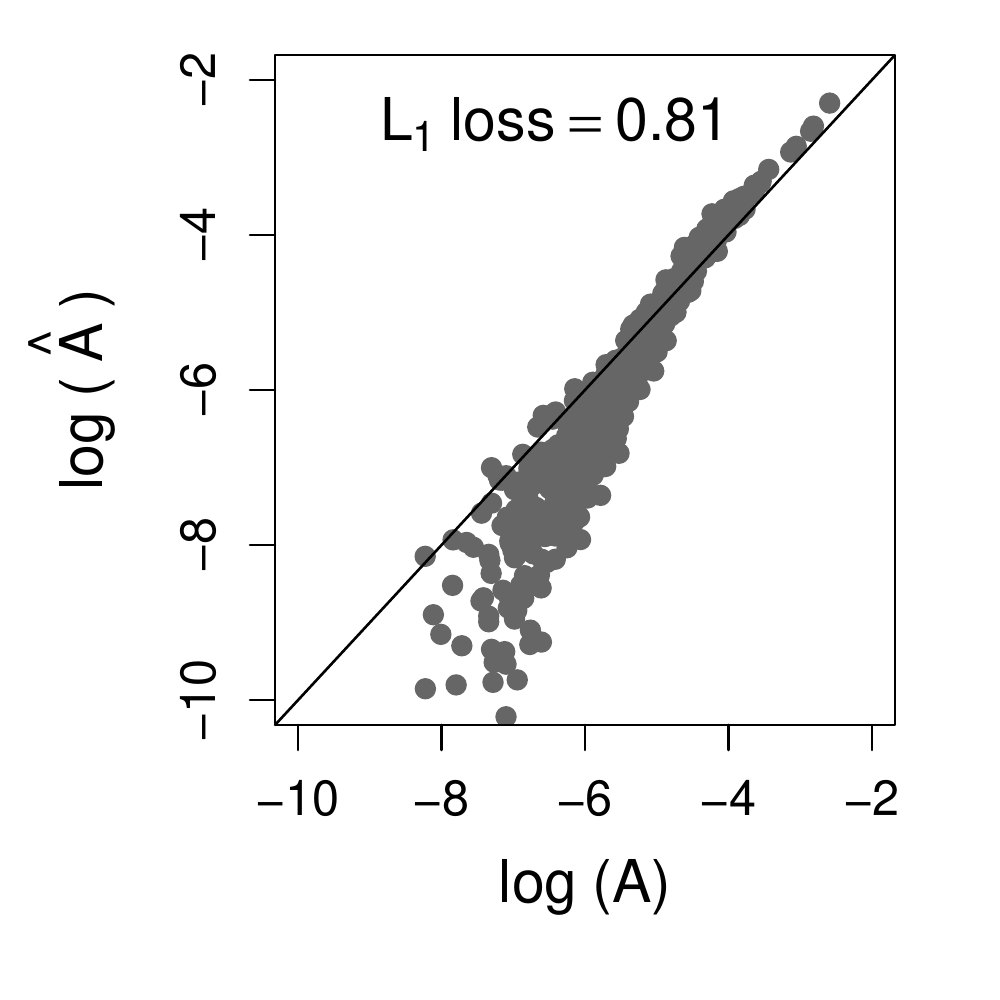}
	\end{subfigure}
		\begin{subfigure}[t]{0.24 \textwidth}
		\centering
		\caption{Submodel}\label{fig:sc-estA}
		\includegraphics[width=\textwidth]{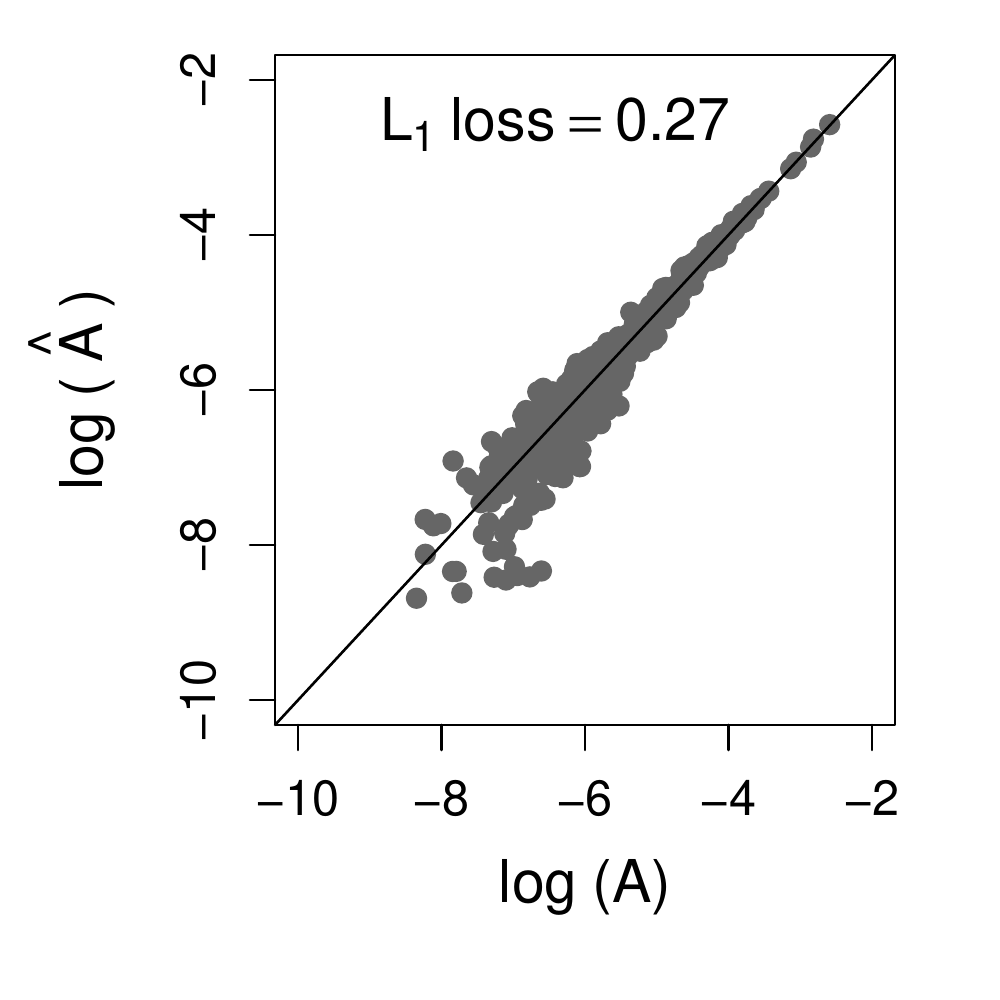}
	\end{subfigure}
	\begin{subfigure}[t]{0.24 \textwidth}
		\centering
		\caption{URSM}\label{fig:unif-estA}
		\includegraphics[width=\textwidth]{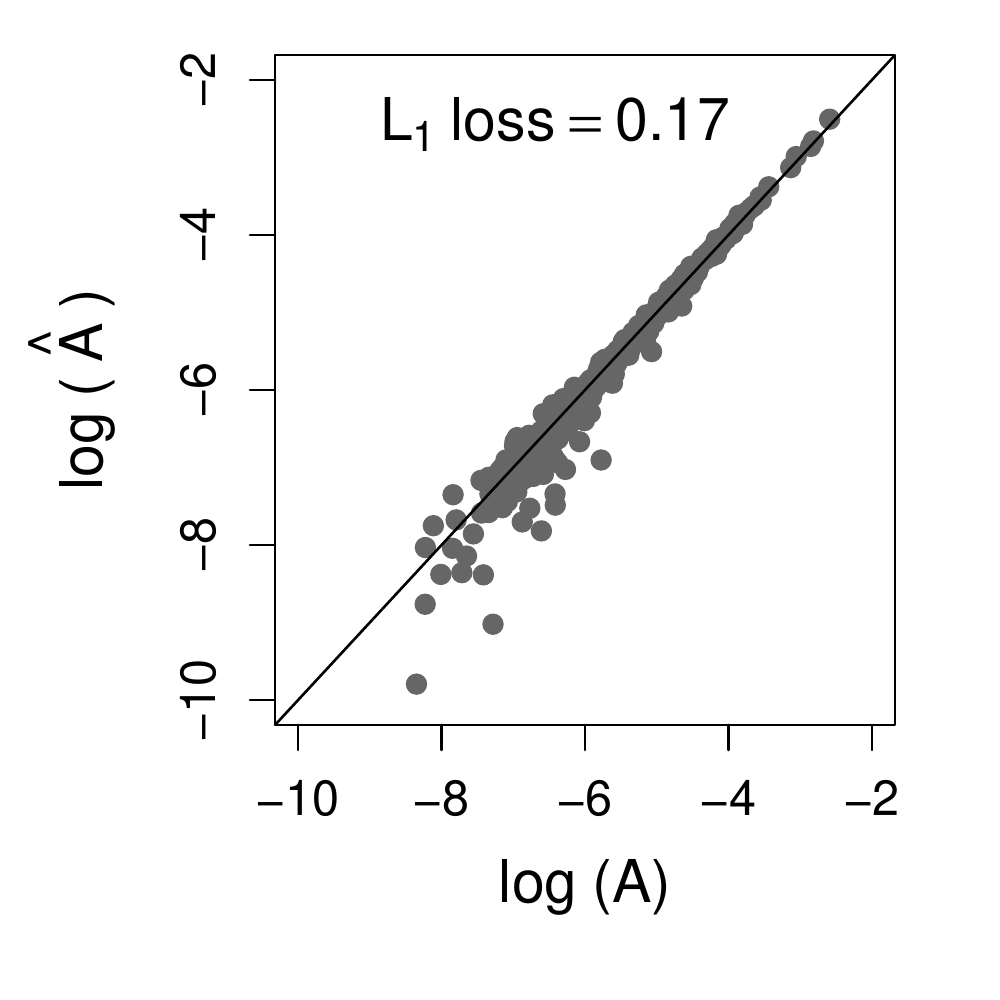}
	\end{subfigure}
	\caption{{\bf (a)} Simulated logistic dropout probability curves for 100 single cells, as defined in equation \cref{eq:dropout-curve}.
	{\bf (b) -- (d)} True profile matrix $A$ versus the estimated $\hat{A}$, plotted in the log scale, using (i) the naive sample mean estimation (equation \cref{eq:samplemean}); (ii) a submodel using only single cell data; (iii) URSM with both single cell and bulk data. The $L_1$ loss $\sum_{i,k} | \hat{A}_{ik} - A_{ik} | $ is reported on the top.}
\end{figure}

In this section, we illustrate that URSM provides accurate estimation on the profile matrix $A$ after correcting for dropouts and utilizing bulk samples.
Following the simulation settings specified in \Cref{sec:sim-setting}, we generate $L=100$ single cells and $M=150$ bulk samples on $N=200$ genes. We consider $K=3$ cell types. \textcolor{highlight}{
For single cells, $30\%, 30\%$ and $40\%$ of the cells are assigned to the 3 different types, respectively. For bulk samples, the hyper parameter of the mixing proportions is set to $\alpha = (1, 2, 3)$. 
The dropout probability curves, simulated following equation \cref{eq:dropout-curve}, are shown in \cref{fig:dropout-curve}. The simulated single cell data has $64.6\%$ entries being zero. 
}

A naive method to estimate the profile matrix $A$ is to use the sample means of single cell expression levels, after normalizing by their sequencing depths. Specifically, recall that $Y \in {\mathbb{R}}^{N \times L}$ represents the observed expression levels in single cells, $\{G_l\}_{l=1, \cdots, L}$ represent the cell types, and $\{ R_l \}_{l=1, \cdots, L}$ are the sequencing depths, defined as $R_l = \sum_{i} Y_{il}$. Then an entry $A_{ik}$ can be estimated by 
\begin{equation} 
\hat{A}_{ik}^{naive} = \frac{1}{ \# \{l: G_l=k\}} \sum_{l: G_l=k} \frac{Y_{il}}{R_l} \,.
\label{eq:samplemean}
\end{equation}
However, due to the presence of dropout events and the dependency between $\pi_{il}$ and $A$, 
\textcolor{highlight}{
this naive sample mean estimation is biased, 
with $L_1$ loss 0.81 (\cref{fig:sample-estA}), where the $L_1$ loss is computed as $\sum_{i,k} | \hat{A}_{ik} - A_{ik} | $.} 
On the other hand, by explicitly modeling the occurrence of dropout events and capturing the relationship between dropout probability and expected expression level, 
\textcolor{highlight}{
a submodel of URSM that only uses single cell data successfully corrects for the bias, and substantially reduces the loss to 0.27 (\cref{fig:sc-estA}). Finally, by integrating the bulk data, URSM further improves the estimation and further reduces the $L_1$ loss to $0.17$ (\cref{fig:unif-estA}). 
}


\subsection{Deconvolution of bulk samples}
Now we further examine the model performance on inferring the mixing proportions $W$ in bulk samples, using the same simulation setting as in \Cref{sec:motivate}. 
We compare the performance of URSM to three widely used deconvolution methods: Digital Sorting Algorithm (DSA)  \citep{zhong2013digital}, semi-supervised Nonnegative Matrix Factorization (ssNMF) \citep{gaujoux2012semi}, and Cibersort \citep{newman2015robust}. 

\begin{figure}	
	\centering
	\begin{subfigure}[t]{0.45 \textwidth}
		\centering
		\caption{Estimating $A$}
		\includegraphics[width=\textwidth]{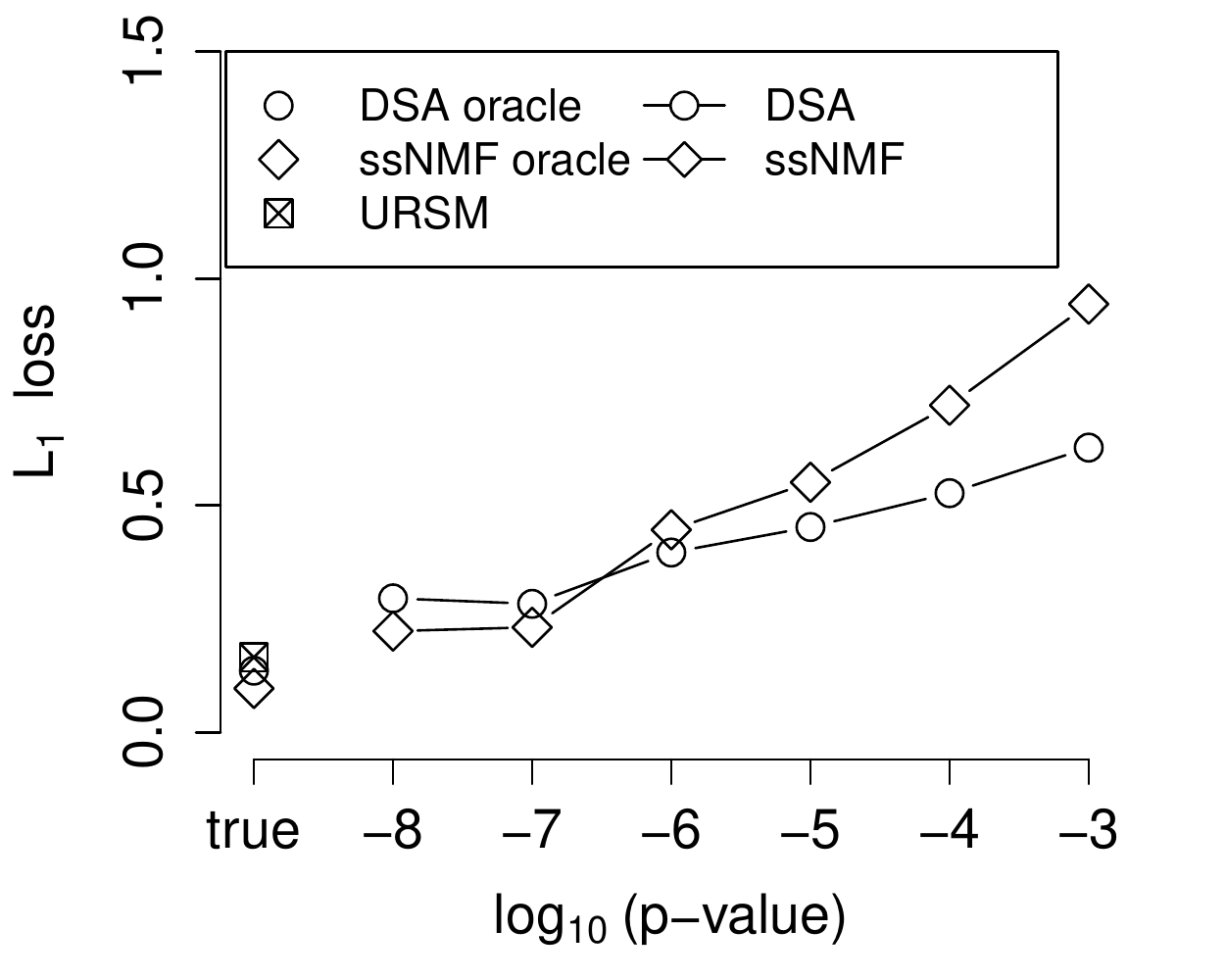}		
	\end{subfigure}
	\quad
	\begin{subfigure}[t]{0.45 \textwidth}
		\centering
		\caption{Estimating $W$}\label{fig:w-l1error}
		\includegraphics[width=\textwidth]{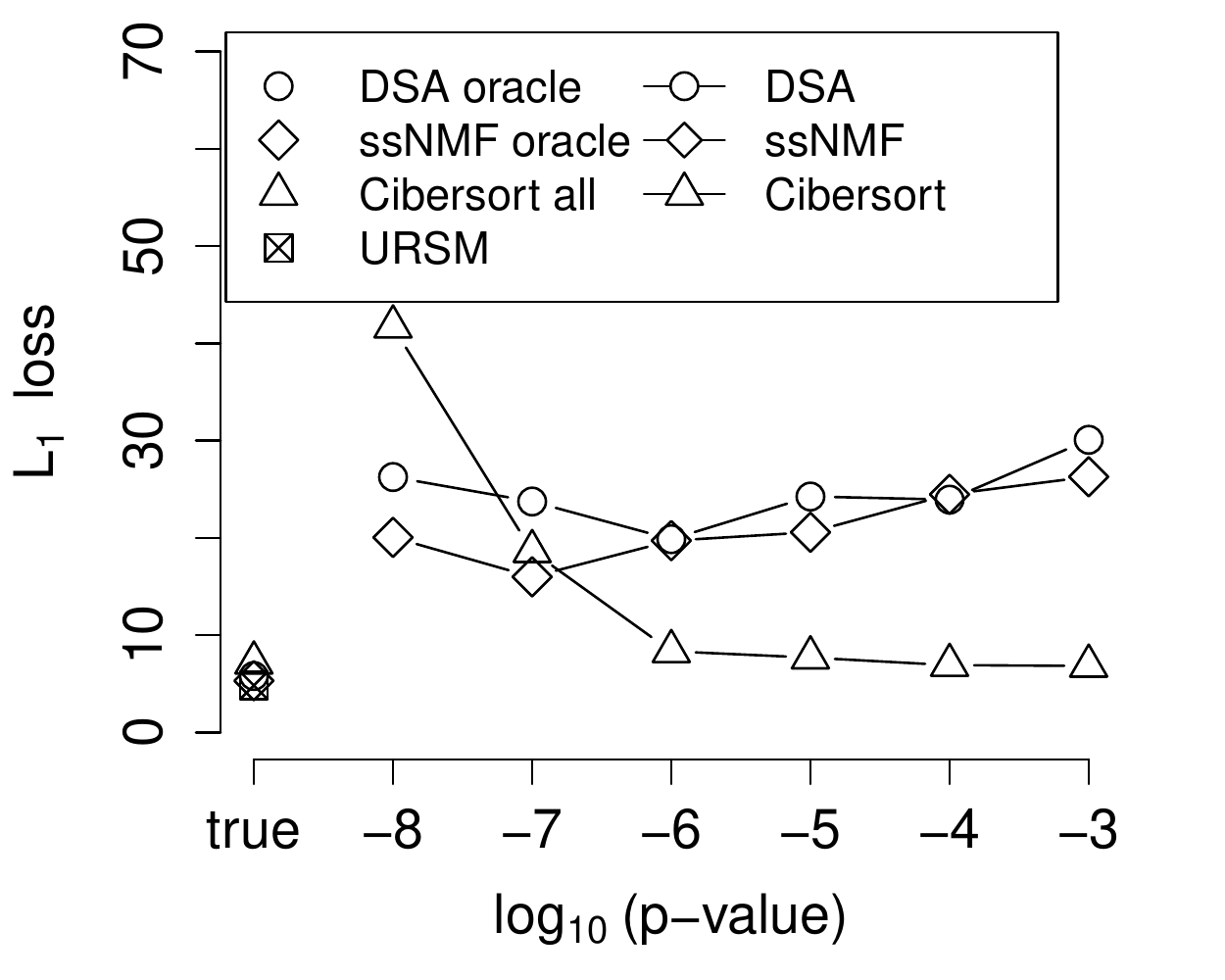}
	\end{subfigure}
	\caption{The $L_1$ loss of recovering {\bf (a)} the profile matrix, $\sum_{i,k} | \hat{A}_{ik} - A_{ik} |$, and {\bf (b)} mixing proportions, $\sum_{k, j} | \hat{W}_{kj} - W_{kj} |$.  We evaluate DSA and ssNMF when the marker genes are extracted from single cell data using different thresholds of $p$-values, as well as under the oracle condition where the true marker genes are given. 
We evaluate Cibersort on estimating $W$ when the input signature matrix is based on the estimated $\hat{A}$ from URSM. We report its performance when the entire $\hat{A}$ is used (``Cibersort all"), as well as when only the estimated marker genes are used (``Cibersort").
The performance of URSM is plotted with a square in both panels, which does not depend on thresholding $p$-values. 
}
	\label{fig:sim-deconv}
\end{figure}

Both DSA and ssNMF rely heavily on a set of {\it given} marker genes as input to guide the matrix factorization, where a ``marker gene" is only expected to express in one cell type. 
Unfortunately, marker genes are rarely known in practice, and a widely adopted procedure is to estimate the list of marker genes from purified cells by selecting those with the most different expression levels across cell types. Here, we mimic this procedure by estimating a list of marker genes from single cell data to guide DSA and ssNMF. Specifically, we adopt  the method in \cite{abbas2009deconvolution}, which calculates a $p$-value of each gene by comparing its expression level in the highest and second-highest types of cells, then selects the group of genes with the smallest $p$-values. \cref{fig:sim-deconv} shows the $L_1$ loss of estimating $A$ and $W$ using DSA and ssNMF with different sets of estimated marker genes with $p$-values smaller than $\{10^{-8}, \cdots, 10^{-3}\}$, and 
\textcolor{highlight}{the number of selected marker genes is listed in \Cref{tab:marker}}. 
It is clear that these two algorithms are sensitive to the input marker genes. For comparison, we also evaluate the performances of DSA and ssNMF  when the oracle information of true marker genes is available.

\textcolor{highlight}{
On the other hand, Cibersort requires a ``signature" matrix containing the expression levels of a group of ``barcode" genes that collectively distinguish between different cell types. Note that this essentially requires knowing part of the profile matrix $A$, which contains much more information than the marker gene list. 
Here, we use the estimated $\hat{A}$ from our unified model as the signature matrix for Cibersort. We report the $L_1$ loss of estimating $W$ when Cibersort only takes the expression levels of the selected marker genes, as well as when Cibersort uses the entire $\hat{A}$. \Cref{fig:w-l1error} suggests that Cibersort prefers larger number of barcode genes as input.
}

Finally, URSM automatically utilizes the information in single cell data to guide deconvolution. \cref{fig:sim-deconv} illustrates that URSM and Cibersort usually outperform DSA and ssNMF using estimated marker genes, and achieve comparable $L_1$ loss even when DSA and ssNMF have the oracle information of marker genes.

\bgroup
\def\arraystretch{1.2} 

\begin{table}[htbp]
\caption{Number of selected marker genes using different thresholding $p$-values. }
\label{tab:marker}
\begin{center}
\begin{tabular}{ r | r r r r r r | r }
\hline
& \multicolumn{6}{c | }{$\log_{10}( \text{p-value}) $} &  True markers  \\
\# of markers  & -8 & -7 & -6 & -5 & -4 & -3  &  \\
\hline
cell type 1 &  5 & 8 & 11 & 16 & 19 &  27 & 10 \\
cell type 2 & 2 & 2 & 8 & 11 & 16 & 23 & 10 \\
cell type 3 & 1 & 2 & 7 & 10 & 17 & 21  & 10 \\
\hline
\end{tabular}
\end{center}
\label{default}
\end{table}%

\subsection{Inference of dropout entries in single cell data}
Next, we present the inference on dropout entries in single cell data, again using the same setting as in \Cref{sec:motivate}. Here, our goal is to distinguish between dropout entries and structural zeros, as defined in equation \cref{eq:drop-struct}. Note that we only need to make inference for locations where the observed expression levels are zero, i.e., on the set $\{(i, l): Y_{il} = 0\}$. Recall that $S_{il}=0$ if gene $i$ is dropped out in cell $l$, and our model provides the estimated posterior mean of $S$:
\begin{equation}
 \widetilde{\pi}_{il} =  \Exp( S_{il} ~|~ X, Y, \theta )\,, 
 \label{eq:tilde-pi}
 \end{equation}
where $\theta$ denotes the model parameters. Hence a natural approach is to predict the entries with small $\widetilde{\pi}_{il}$ to be dropouts.

A potential competitor for imputing dropout entries is the Nonnegative Matrix Factorization (NMF) \citep{lee2001algorithms}. One can construct a low-rank approximation to the single cell expression matrix $Y \in \mathbb{R}^{N \times L}$ using NMF.  Intuitively, the approximated values tend to be higher at dropout entries, and closer to zero at structural-zero entries. 
\textcolor{highlight}{
As shown in \cref{fig:roc},  if the rank is properly specified, this simple NMF-based method demonstrates certain ability to distinguish between dropout genes and structural zeros, but not as well as URSM.
}
In addition, in order to further impute the dropout entries, a good estimation of the profile matrix $A$ is also needed.  \Cref{fig:nmf-estA} shows the estimation of $A$ by taking sample average as in equation \cref{eq:samplemean}, with $Y$ substituted by the NMF approximation. It is clear that the NMF approach fails to correct for the bias introduced by the dropout events, while URSM succeeds in both identifying dropout entries and obtaining an unbiased estimation of $A$ (recall \cref{fig:unif-estA}).

\begin{figure}	
	\centering
	\begin{subfigure}[t]{0.4 \textwidth}
		\centering
		\caption{ROC curves}\label{fig:roc}
		\includegraphics[width=\textwidth]{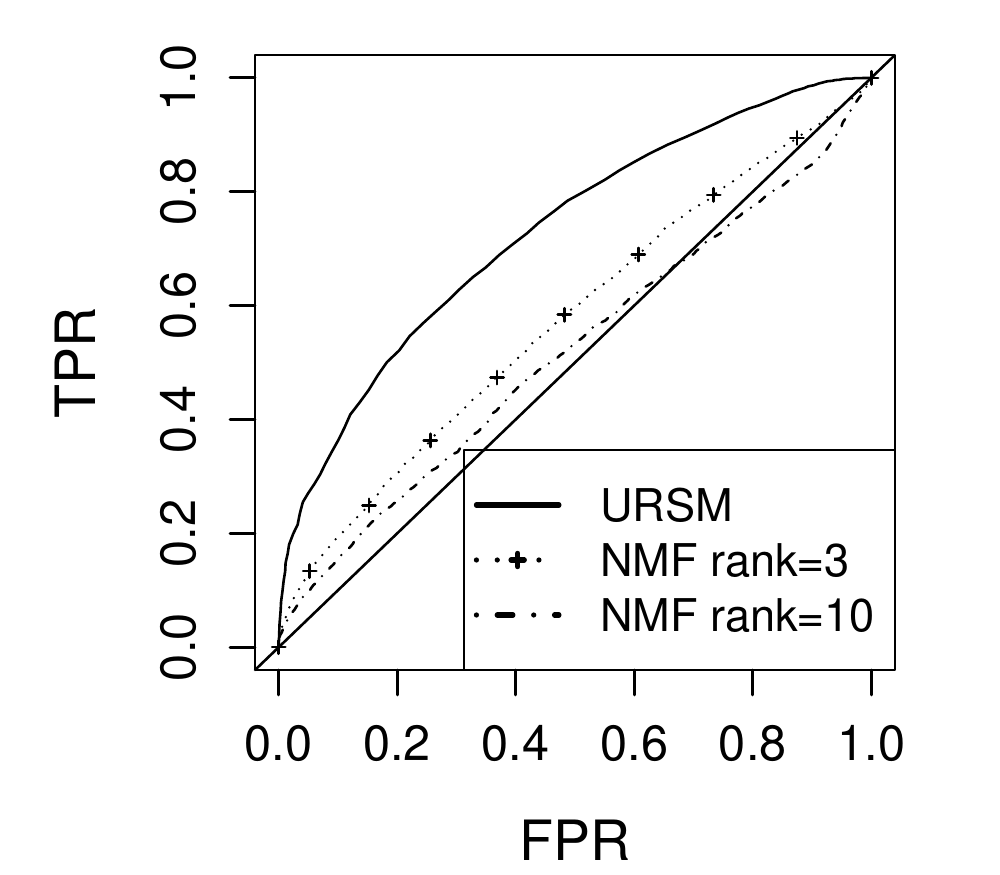}		
	\end{subfigure}
	\quad
	\begin{subfigure}[t]{0.4 \textwidth}
		\centering
		\caption{NMF mean estimation}\label{fig:nmf-estA}
		\includegraphics[width=\textwidth]{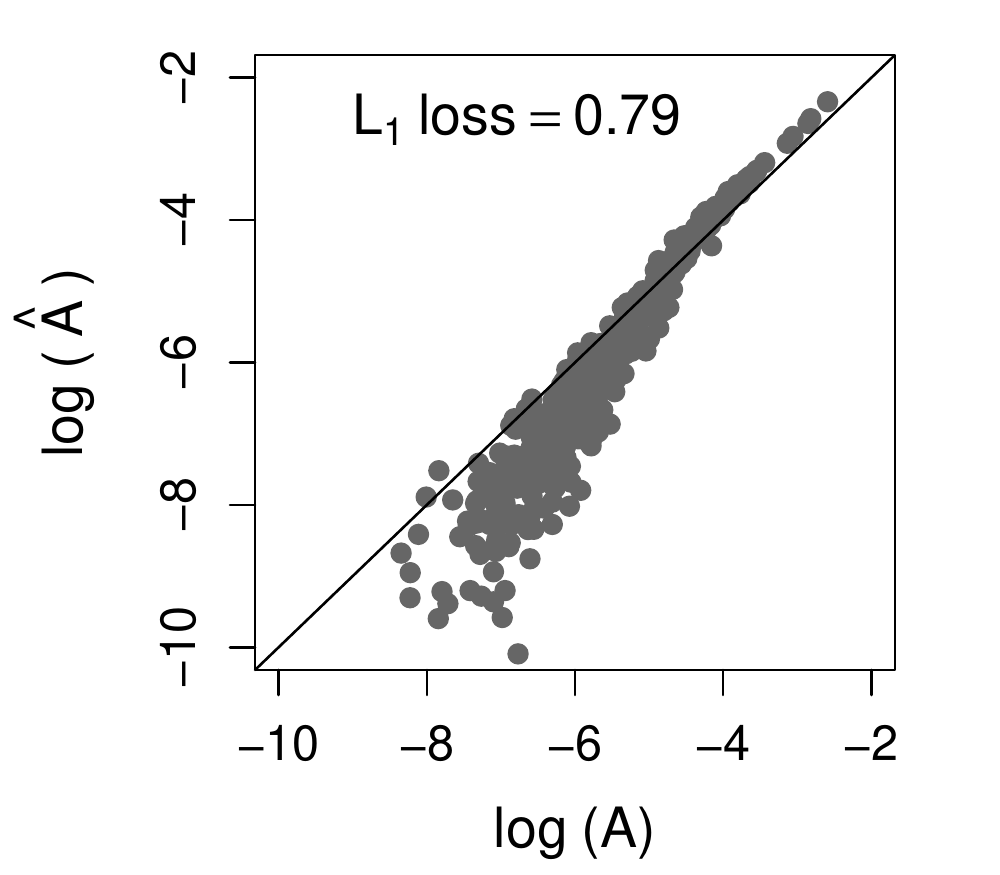}
	\end{subfigure}
	\caption{
	{\bf (a)} ROC curves of identifying dropout entries in single cell data. 
	{\bf (b)} True profile matrix $A$ versus the sample average of a rank-3 NMF approximation, plotted in the log scale. The $L_1$ loss $\sum_{i,k} | \hat{A}_{ik} - A_{ik} |$ is reported on the top.}
\end{figure}

\subsection{Robustness}
\textcolor{highlight}{
Finally, we demonstrate the robustness of our model. We apply URSM under the scenario where the number of cell types in single cell data $K_{sc}$ is not equal to the number of cell types in bulk data $K_{bk}$, as well as when the number of genes $N$ grows. URSM always takes $K = \max\{K_{sc}, K_{bk}\}$ as input, and estimates $\hat{A}_{unif} \in \mathbb{R}^{N \times K}$ and $\hat{W}_{unif} \in \mathbb{R}^{K \times M}$. When $K_{sc} > K_{bk}$, 
it is straightforward to directly apply URSM, and 
ideally the estimated $\hat{W}_{unif}$ will assign zero proportions to the missing cell types in bulk samples. However, when  $K_{ sc} < K_{bk}$, 
without extra information, deconvolution is an ill-defined problem because of the non-identifiability issue (see \Cref{sec:introduction} for more details). 
In order to  find a biological meaningful solution, we follow the idea in ssNMF \citep{gaujoux2012semi} and use a set of marker genes to initialize the parameters for the EM algorithm by setting the corresponding entries in $A$ to be zero. 
We consider the scenario where for each cell type, 5 true marker genes and 3 imperfect marker genes are used for initialization. 
The imperfect marker genes are selected from the non-marker genes, where we pick the ones with the largest difference between the highest and second highest expression levels across cell types in $A$.
}

\textcolor{highlight}{
Following \Cref{sec:sim-setting}, we simulate $M=150$ bulk samples, where the mixing proportions in bulk samples are generated from Dir$(\alpha)$ with $\alpha = (1, ..., K_{bk})$. For single cell data, we generate $40$ cells in the majority cell type, and $30$ cells in each of the remaining $K_{sc}-1$ types. To reduce the computation load and enhance stability, we use the maximum a posteriori estimation for $W$ in the E-step for bulk samples. More details are included in the supplement.   
}

\textcolor{highlight}{
Again, we compare URSM to  DSA, ssNMF, and Cibersort. Both DSA and ssNMF require a set of marker genes as input, and we report their performances under two scenarios: (i)  the oracle scenario where 5 true marker genes are provided for each cell type; and (ii) a more realistic scenario as used by our uniform model, where 5 true marker genes and 3 imperfect marker genes are provided for each cell type. Note that when $K_{sc} > K_{bk}$, bulk samples contain no information of the expression patterns for the missing cell types, so we allow DSA and ssNMF to only deconvolve $K_{bk}$ cell types in these cases. We point out that this strategy favors the DSA and ssNMF methods by providing them extra information of the missing cell types in bulk samples. For Cibersort, as in the previous sections, we use  the estimated profile matrix obtained from our uniform model as the input signature matrix.
}

\textcolor{highlight}{
\Cref{fig:robust} summarizes the performance of different models under various choices of $K_{sc}$ and $K_{bk}$ when $N=200$ in 10 repetitions. In order to make a comparable comparison across different $K$'s, we report the  {\it average per cell type} $L_1$ loss, i.e., the average $L_1$ loss 
$|| \hat{A}_{\cdot, k} - A_{\cdot, k} ||_1$ and $|| \hat{W}_{\cdot, k} - W_{\cdot, k} ||_1 $ across all columns $k$. We see that the performance  of URSM remains robust under different settings, and is usually comparable to DSA and ssNMF algorithms  even when the latter two algorithms have the oracle marker gene information. Not surprisingly, Cibersort has similar performance as URSM because it uses our estimated $\hat{A}_{unif}$ as input.  We point out that when the sample mean estimation $\hat{A}_{naive}$ (equation \cref{eq:samplemean}) is given to Cibersort as the signature matrix, the performance is unstable and it cannot provide deconvolution when $K_{sc} < K_{bk}$.
Finally, we also demonstrate the performance of different models when $N=\{200, 500, 1000\}$, where we set $K_{sc}=K_{bk}=3$. \Cref{fig:scale} verifies that URSM remains robust with larger numbers of genes.
}

\begin{figure}	
	\centering
	\begin{subfigure}[t]{ \textwidth}
		\centering
		\includegraphics[width=\textwidth]{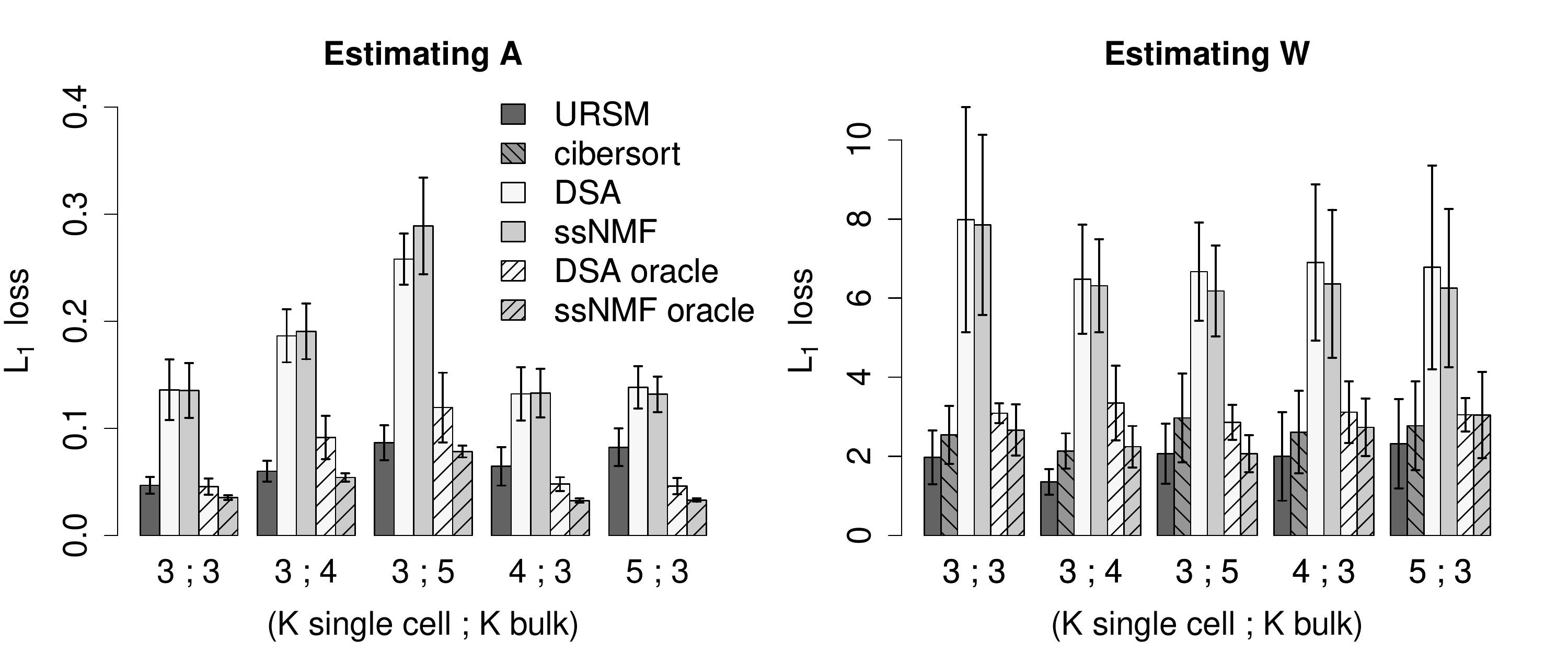}
		\caption{$N=200$, varying $K_{sc}$ and $K_{bk}$}	
		\label{fig:robust}	
	\end{subfigure}
	\begin{subfigure}[t]{ \textwidth}
		\centering
		\includegraphics[width=\textwidth]{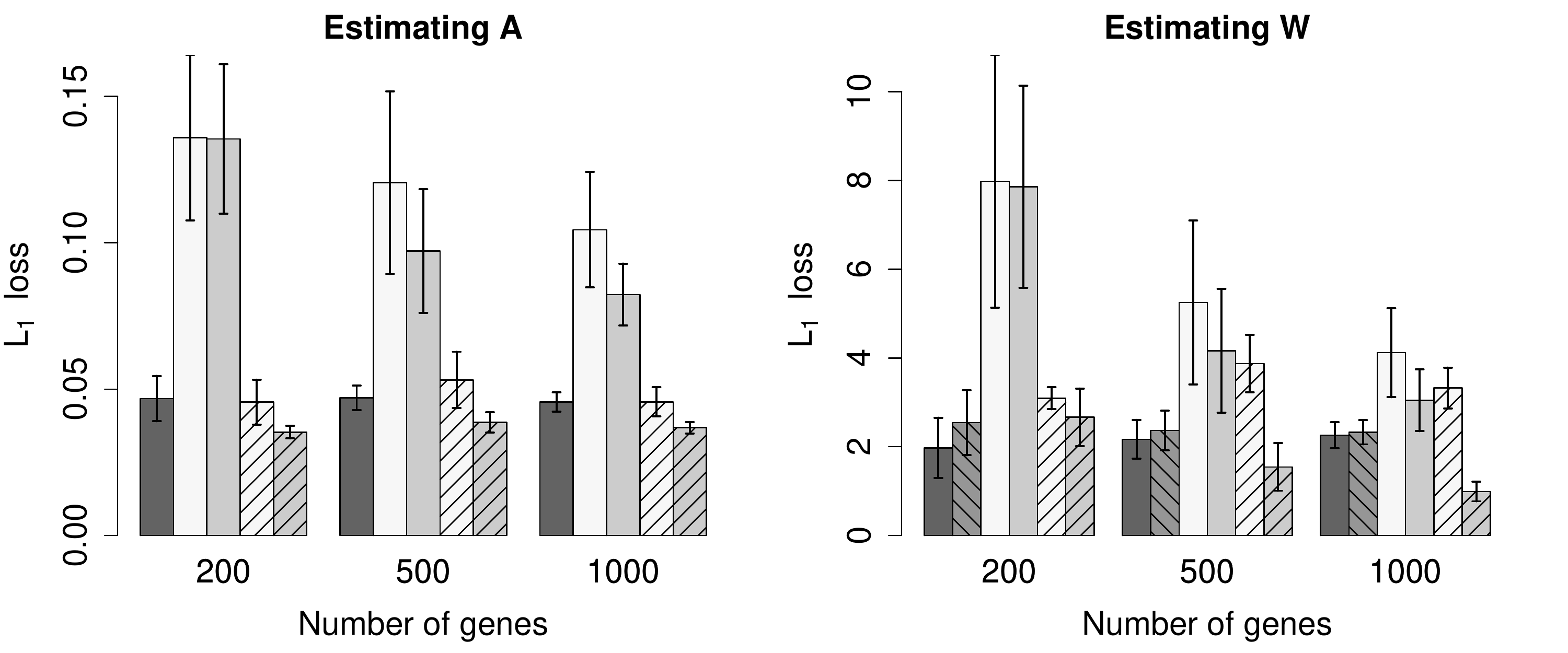}
		\caption{ $K_{sc} = K_{bk} =3$, varying $N$} \label{fig:scale}		
	\end{subfigure}
	\caption{The average per cell type $L_1$ loss of recovering the profile matrix $A$ and the mixing proportions $W$ in 10 repetitions, with the standard deviations shown by the error bars, when {\bf (a)} $K_{sc}, K_{bk} \in \{3,4,5\}$ with $N=200$ genes; {\bf (b)} $N=\{200, 500, 1000\}$ with $K_{sc} = K_{bk} = 3$. Each figure shows the performance of (i) URSM; (ii) DSA and ssNMF with 5 true marker genes and 3 imperfect marker genes per cell type as input; (iii) DSA and ssNMF under the oracle scenario where 5 true marker genes per cell type are provided. We also report the performance of Cibersort for estimating $W$ using the estimated $\hat{A}_{unif}$ from URSM as the input signature matrix. }
\end{figure}


\section{Application to Fetal Brain Data}
\label{sec:data}

\subsection{Data pre-processing}
In this section, we apply URSM to gene expression measured on fetal brains. The single cell RNA-seq data comes from \cite{camp2015human}, where 226 cells from fetal brains are sequenced on 18,927 genes. The authors have removed endothelial cells and interneurons, and the remaining 220 single cells are labeled into three types: 40 apical progenitors (APs), 19 basal progenitors (BPs), and 161 neurons (Ns). 
\textcolor{highlight}{
In addition, the authors have normalized the RNA-seq counts using FPKM (Fragments Per Kilobase of exon per Million fragments mapped) and performed log-transformation by $\log_2(x+1)$. We refer the readers to \cite{camp2015human} for more details of the single cell data pre-processing.}
 On the other hand, microarray bulk gene expression data on fetal brains is provided by the BrainSpan atlas \citep{kang2011spatio}. Within the same window of development, 12 to 13 post-conception week, 72 bulk samples from prefrontal cortex  are measured on 16,947 genes. To apply our model, the single cell RNA-seq data are transformed back to linear scale by $2^x - 1$, and all measurements are truncated to integers. To approximate the RNA-seq counts in bulk samples, we transform the BrainSpan microarray data in the same way and treat them as pseudo-RNA-seq counts.  The resulting bulk samples have an average pseudo sequencing depth of $5.5 \times 10^6$, which is 26 times larger than the average effective sequencing depth in single cells, $2.1 \times 10^5$, where the effective sequencing depth is calculated as the sum of FPKM across all genes in each single cell.

To reduce computational load, we only focus on genes with significantly different expression levels among the three cell types. Specifically, we use the 315 so-called PC genes proposed in \cite{camp2015human}, which have the largest loadings in a Principal Component Analysis (PCA) and account for the majority of cellular variation. After restricting to the overlapping genes that are also sequenced in BrainSpan bulk samples, a final list of 273 PC genes are obtained and used in the following analyses. When restricting to these 273 genes, the average effective sequencing depth (i.e., the sum of RNA-seq counts in each sample) is 
\textcolor{highlight}{
$3.2 \times 10^5 (sd = 1.6 \times 10^4)$
} in BrainSpan tissues, and 
\textcolor{highlight}{
$1.4 \times 10^4 (sd = 4.3 \times 10^3)$
} in single cells .

Due to the nature of active cell development from APs and BPs to Neurons in fetal brains, we expect to have a few cells that are actively transitioning between two cell types, whose labels are ambiguous. We first remove these ambiguously labeled cells from our analysis. Specifically, we project the single cells to the leading 2-dimensional principal subspace, where the pseudo developing time is constructed using the Monocle algorithm \citep{trapnell2014dynamics}. Based on the results, the 3 BPs that are close to AP or Neuron clusters are removed, so are the 4 Neurons that are close to AP or BP clusters (\Cref{fig:monocle}). The remaining 213 single cells are retained for analysis, and their gene expression levels on the 273 PC genes are visualized in \Cref{fig:scdata}.

\subsection{Imputation of single cell data}
Here, we apply URSM to identify and impute the dropout entries in single cell data. Note that in order to distinguish between dropout entries and structural zeros in single cell data (equation \cref{eq:drop-struct}), we only need to focus on the entries where the observed gene expression levels are zero. 
The inference of dropout entries is based on the estimated posterior expectation of $\Exp( S_{il} ~|~ X, Y, \theta )$. 
As a result, among the 37,771 zero-observation entries, 45.7\% are inferred to be dropouts with probability one (\cref{fig:impute}). These entries are then imputed by their expected values, calculated using the corresponding entries in the estimated profile matrix $A$ multiplied by the sequencing depths of the corresponding cells. To illustrate the impact of imputation, we apply PCA again on the imputed data. \cref{fig:impute-pca} visualizes the cells in the first two principal components, and the clusters for different cell types are more clearly separated. 
 
\begin{figure}	
	\centering	
	\begin{subfigure}[t]{0.45 \textwidth}
		\centering
		\caption{Cleaned single cell data} \label{fig:scdata}	
		\vspace{-7pt}	
		\includegraphics[width=\textwidth]{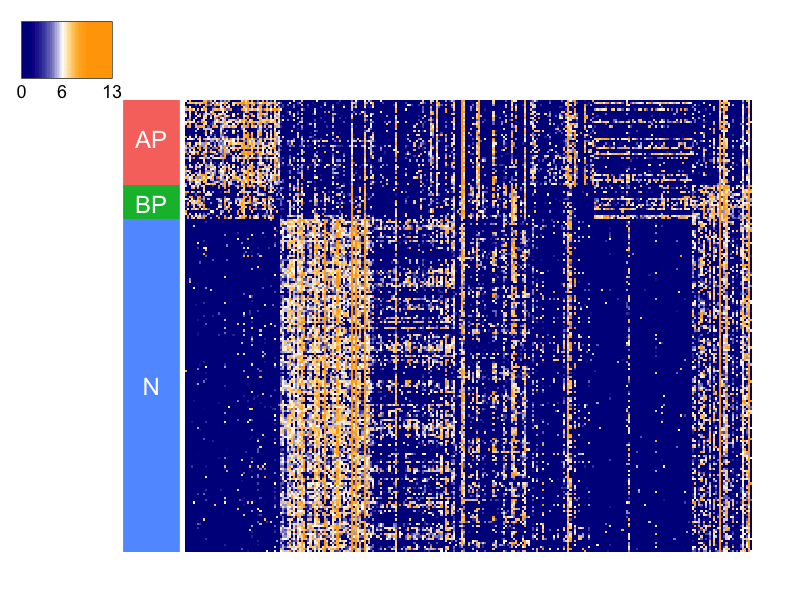}		
	\end{subfigure}
	\quad
	\begin{subfigure}[t]{0.45 \textwidth}
		\centering
		\caption{PCA on original data}\label{fig:monocle}		
		\includegraphics[width=\textwidth]{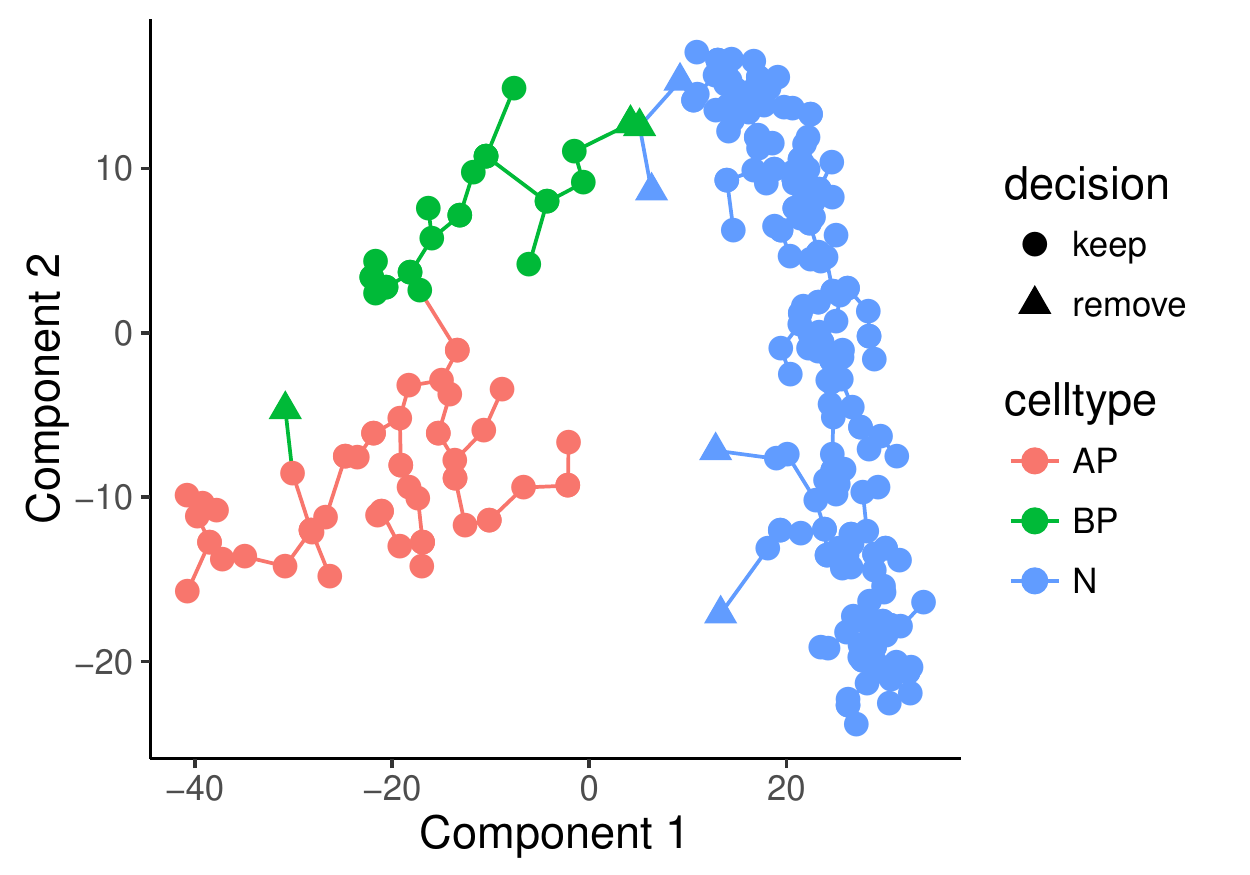}	
		\vspace{3pt}	
	\end{subfigure}
	
	\vspace{5pt}
	\begin{subfigure}[t]{0.45 \textwidth}
		\centering
		\caption{Imputed entries} \label{fig:impute}
		\vspace{-7pt}	
		\includegraphics[width=\textwidth]{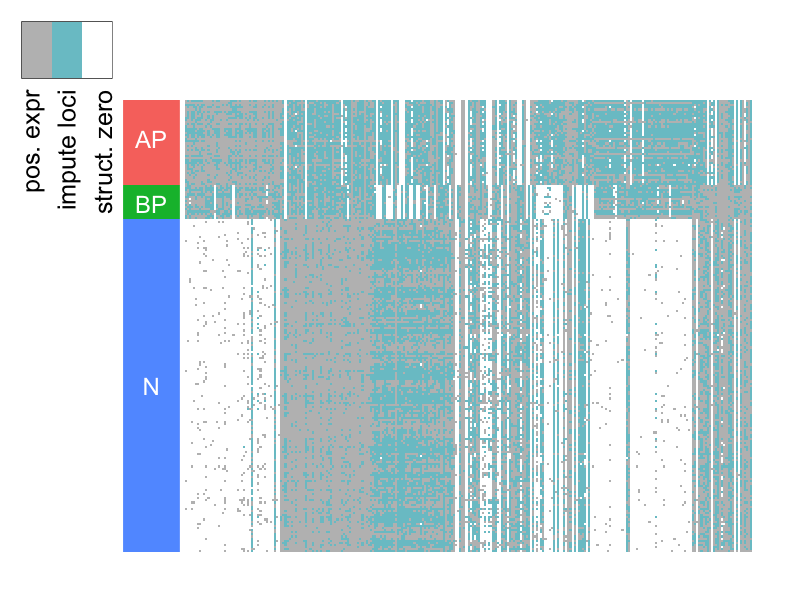}		
	\end{subfigure}
	\quad
	\begin{subfigure}[t]{0.45 \textwidth}
		\centering
		\caption{PCA on imputed data}\label{fig:impute-pca}
		\includegraphics[width=\textwidth]{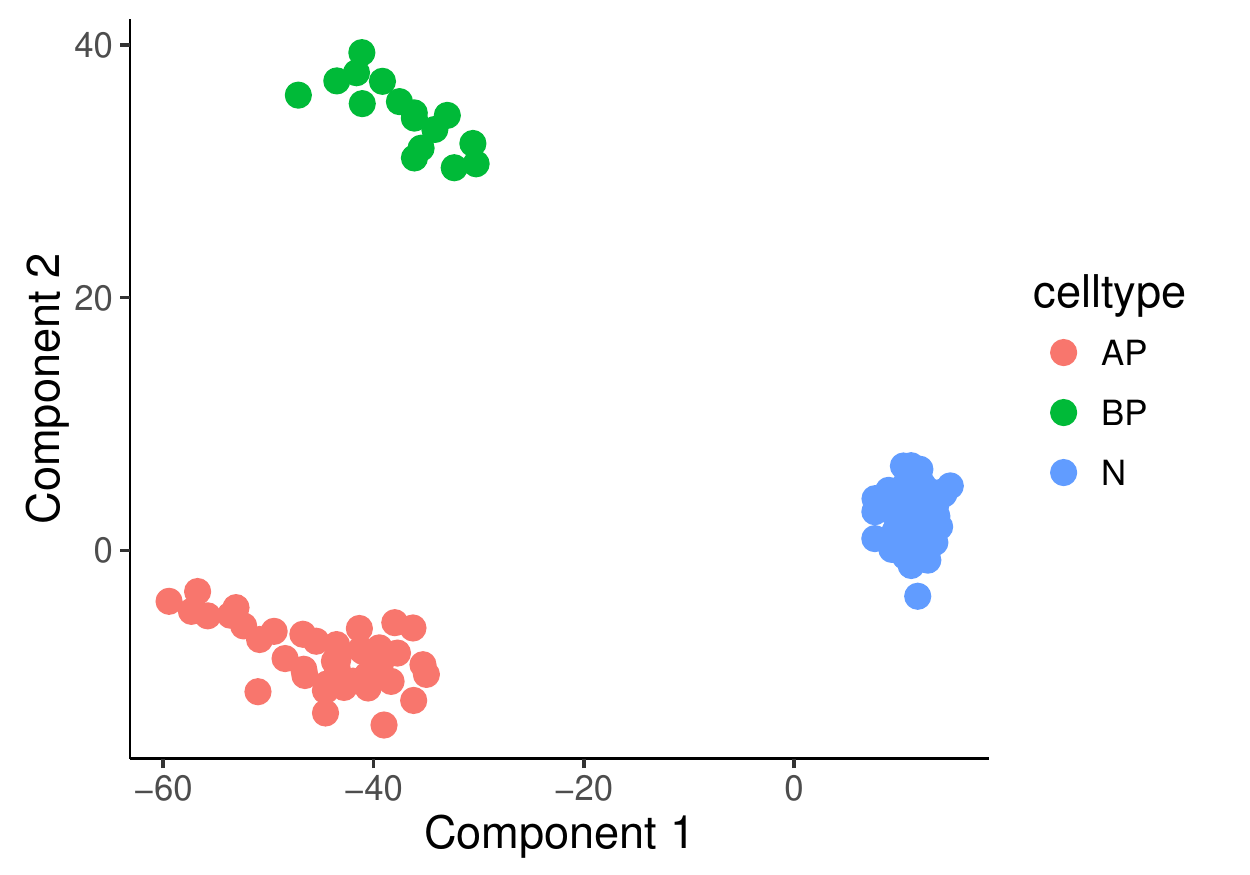}		
	\end{subfigure}

	\caption{
	{\bf (a)} Single cell gene expressions ($\log_2$(FPKM+1)) after removing 7 ambiguously labeled cells. Rows are 213 cells and columns are 273 genes.
	{\bf (b)} PCA applied on the original single cell data with 220 labeled cells using 273 PC genes, where the Monocle algorithm is applied to construct pseudo developmental times. 7 cells are identified to be ambiguously labeled and are removed from our analyses (marked as triangles).
 	{\bf (c)} Entries in cleaned single cell data that are inferred to be dropout and imputed (marked in blue) versus the entries that are inferred to be structural zeros (marked in white) in cleaned single cell data. The entries with positive expression levels have no need for posterior inference, and are marked in grey.
	{\bf (d)} After imputing dropout genes, PCA is conducted on the 213 cells using 273 PC genes, and the three different types of cells are more clearly separated. }
	\label{fig:pca}
\end{figure}

\subsection{Deconvolution of bulk samples}
Finally, we present the deconvolution results of bulk samples using URSM. According to the prior knowledge that the proportions in bulk samples should be roughly consistent with that in single cell data, the mixing parameter $\alpha$ is initialized at $(2 \times 10^4,\, 10^4,\, 7 \times 10^4)$ for AP, BP and Neurons. The scale of $\alpha$ is chosen to be comparable to the average effective sequencing depths of $1.4 \times 10^4$ among all single cells. \Cref{fig:deconv-unified} shows the inferred mixing proportions of APs, BPs and Neurons in each of the 72 bulk samples, with an average of 17.7\% AP cells, 8.7\% BP cells and 73.6\% Neurons.

For comparison, we also apply the Digital Sorting Algorithm (DSA)  \citep{zhong2013digital}, semi-supervised Nonnegative Matrix Factorization (ssNMF) \citep{gaujoux2012semi}, and Cibersort \citep{newman2015robust}
on the BrainSpan bulk samples. The marker genes for DSA and ssNMF are selected by comparing each gene's expression level in the highest and second-highest types of cells in the single cell data, and genes with $p$-value $<10^{-5}$ are treated as markers \citep{abbas2009deconvolution}. This procedure leads to 21 AP markers, 6 BP markers and 28 Neuron markers, which serve as input to DSA and ssNMF. 
For Cibersort, the input signature matrix is provided by the estimated $\hat{A}$ from URSM.
\Crefrange{fig:sskl}{fig:ciber}  suggest that the proportions estimated by ssNMF tend to have too large variations, while DSA overestimates the neural composition, 
and Cibersort obtains similar results as URSM. 

\begin{figure}	
	\centering
	\begin{subfigure}[t]{0.24 \textwidth}
		\centering
		\caption{URSM}\label{fig:deconv-unified}
		\includegraphics[width=\textwidth]{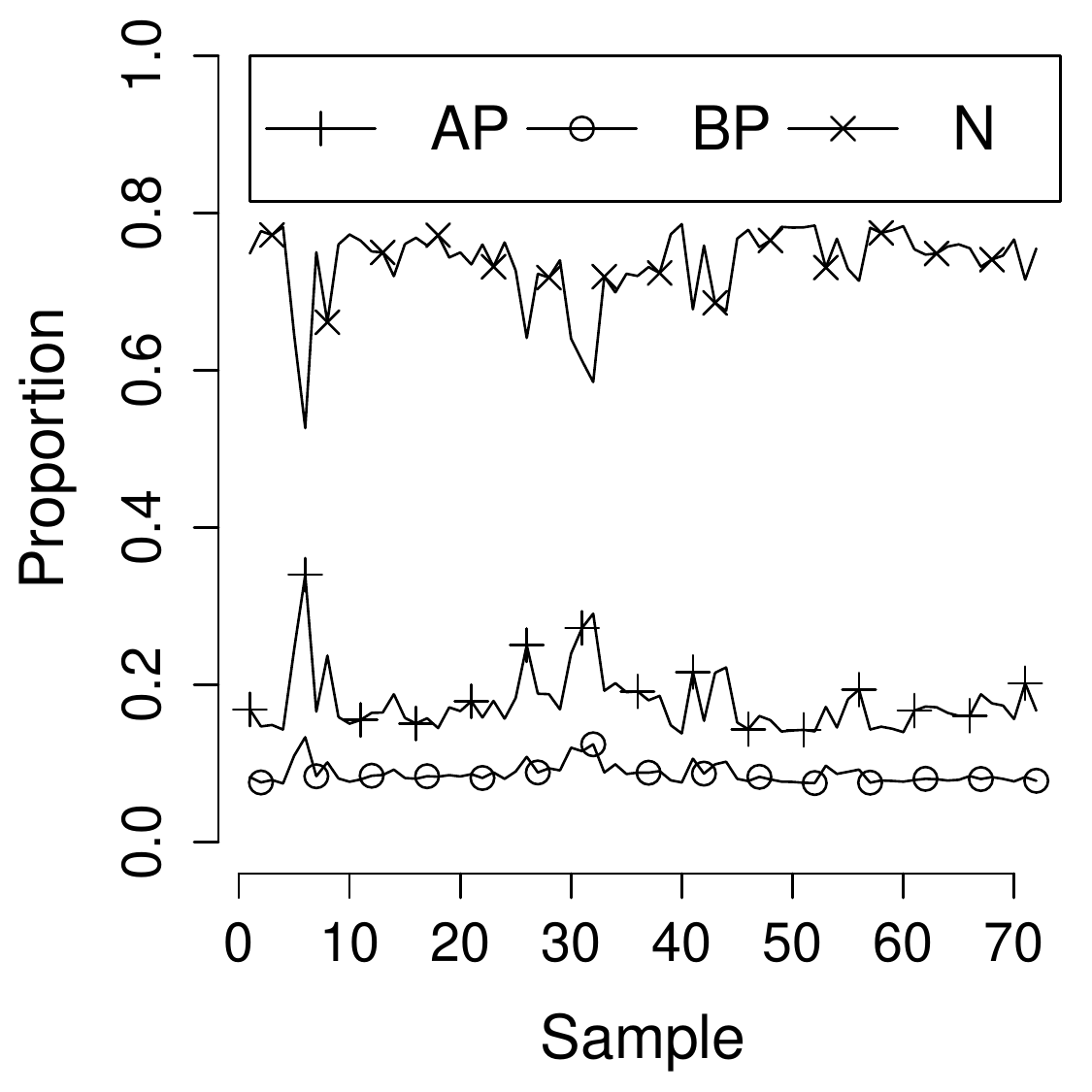}
	\end{subfigure}
	\begin{subfigure}[t]{0.24 \textwidth}
		\centering
		\caption{Cibersort}\label{fig:ciber}
		\includegraphics[width=\textwidth]{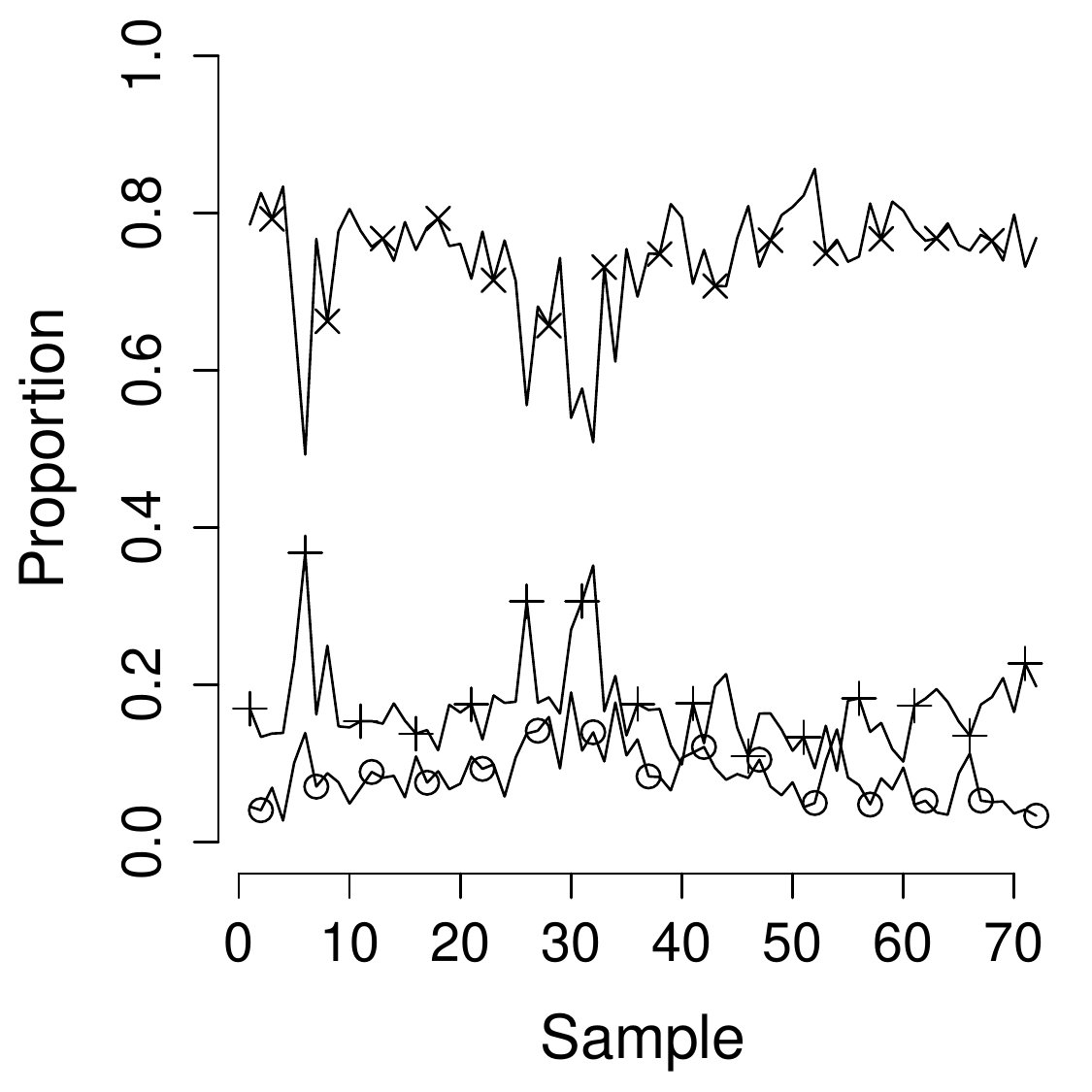}	
	\end{subfigure}
	\begin{subfigure}[t]{0.24 \textwidth}
		\centering
		\caption{DSA}\label{fig:dsa}
		\includegraphics[width=\textwidth]{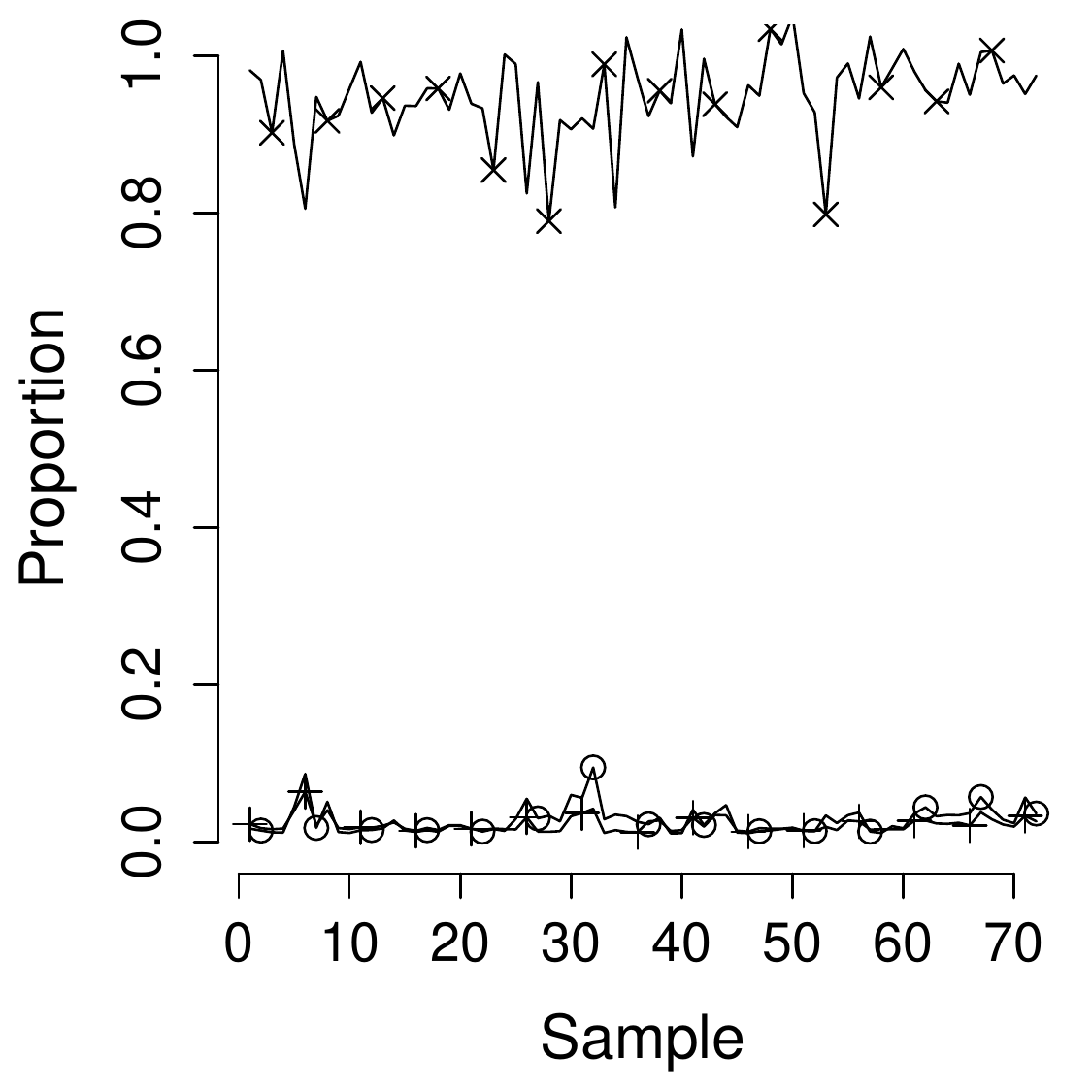}
		
	\end{subfigure}
	\begin{subfigure}[t]{0.24 \textwidth}
		\centering
		\caption{ssNMF}\label{fig:sskl}		
		\includegraphics[width=\textwidth]{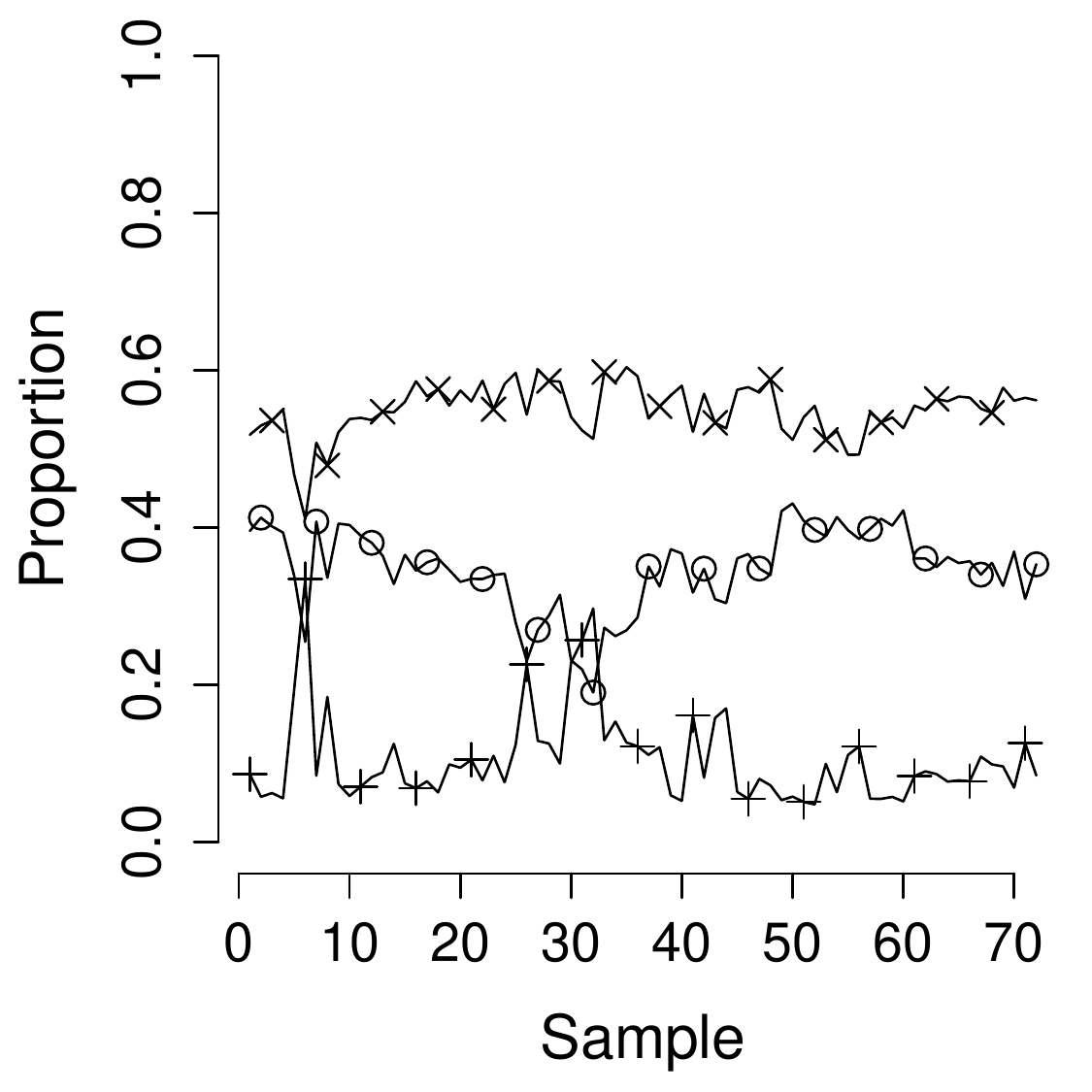}	
	\end{subfigure}

	\caption{Deconvolution of bulk samples into three cell types, using {\bf (a)} URSM; {\bf (b)} Cibersort; {\bf (c)} Digital Sorting Algorithm (DSA);  {\bf (d)} semi-supervised Nonnegative Matrix Factorization (ssNMF).}
	\label{fig:deconv}
\end{figure}

As another perspective to verify the deconvolution results, we use the intuition that the true proportions of a cell type should be correlated with the expression levels of its marker genes in bulk samples. To check whether this holds in the results, we first normalize each bulk sample by their effective sequencing depths, such that the normalized expressions sum to one in each sample. We focus on 7 genes based on biological knowledge, including the radial glia (RG) markers {\it PAX6} and {\it GLI3} that are expected to only express in AP and BP cells, the RG marker {\it HES1} that is mostly expressed in AP cells, the early BP marker {\it HES6}, as well as neuronal genes {\it NEUROD6}, {\it BCL11B} and {\it MYT1L} \citep{camp2015human}. \Cref{tab:corr-marker} summarizes the correlations calculated by estimated proportions using different methods, and we see that URSM and Cibersort usually achieve the highest correlations.
Finally, we point out that if Cibersort uses the naive sample mean estimation from single cell data as the signature matrix, it will fail to identify BP cells and achieve much lower correlations.

\begin{table}[htbp]
\centering
\caption{Correlation between the estimated proportions of a cell type $k$ in bulk samples, $(W_{kj})_j$, and the normalized expression levels $(X_{ij}/R_j)_j$ of its marker gene $i$ in bulk samples. For genes marking both AP and BP, the sum of proportions is used. }
\label{tab:corr-marker}
\begin{tabular}{r r | c c c c}
  \hline
Gene & Marked cell type & URSM & Cibersort & DSA & ssNMF  \\ 
  \hline
HES1 & AP & 0.73 & 0.62 & {\bf 0.80} & 0.68 \\ 
  HES6 & BP & {\bf 0.66} & 0.58 & 0.53 & -0.72 \\ 
  PAX6 & AP.BP & {\bf 0.91} & 0.80 & 0.80 & 0.61 \\ 
  GLI3 & AP.BP & {\bf 0.90} & 0.80 & 0.83 & 0.54 \\ 
  NEUROD6 & N & 0.28 & {\bf 0.37} & 0.02 & -0.36 \\ 
  BCL11B & N & 0.45 & {\bf 0.57} & 0.23 & 0.02 \\ 
  MYT1L & N & {\bf 0.44} & 0.37 & 0.32 & 0.80 \\ 
   \hline
\end{tabular}
\end{table}


\section{Discussion}
\label{sec:discussion}

In this paper, we propose URSM, a unified framework to jointly analyze two types of RNA-seq data: the single cell data and the bulk data. URSM utilizes the strengths from both data sources, provides a more accurate estimation of cell type specific gene expression profiles, and successfully corrects for the technical noise of dropout events in single cell data. As a side product, URSM also achieves deconvolution of bulk data by automatically incorporating the cellular gene expression patterns. 

Dropouts present one of the biggest challenges to modeling scRNA-seq data.   URSM assumes a dependency between expression level and the probability of observing dropout and aims, probabilistically, to infer which observations are likely dropouts. There are a number of alternative approaches in the literature; for a discussion see \cite{huang2017gene}  and \cite{vallejos2017normalizing}.  The most common statistical approach is to explicitly model the zero-inflation process, for example, SCDE \citep{kharchenko2014bayesian}, MAST \citep{finak2015mast:} and ZIFA \citep{pierson2015zifa}.  Some methods assess the fraction of dropouts per gene, other methods, such as CIDR \citep{lin2017cidr:}, take this process to the next step by imputing the dropout values.  SAVER \citep{huang2017gene} avoids trying to determine which observations are dropouts and aims to impute any poorly measured value using the gene-to-gene correlation pattern, and other features in the cell-type specific samples.   

We apply URSM to two gene expression data sets from fetal brains, and obtain promising results on imputing single cell RNA-seq data and deconvolving bulk samples. With more upcoming single cell data on fetal brains, it would be of great scientific interest to apply URSM to specimen from different brain developing periods, which will aid our understanding on gene expression patterns during early brain development and their impact on many complex human disorders. In practice, the degrees of heterogeneity can vary for different tissues. For example, liver tissues may contain more homogeneous cell types. In all cases, URSM can be applied to obtain an accurate estimate of the cell type specific profile.

There are many existing bulk RNA-seq data sets for various human and non-human tissues that can be paired with different single cell data and jointly modeled using this unified framework. We also conduct simulation studies to demonstrate that as long as most cell types are consistent across the two data sources, URSM is robust to subtle mis-matched cell types. 

As for computation, the bottleneck is the Gibbs sampling step, which scales linearly with $N, M, L$ and $K$.  In practice, we find that a few hundred Gibbs samples and 50 -100 EM iterations are usually enough to obtain sensible results. In our experiment, for 100 single cells and 150 bulk samples, one EM iteration with 150 Gibbs samples takes about 3 minutes for 200 genes and 12 minutes for 1,000 genes using a single core on a computer equipped with an AMD Opteron(tm) Processor 6320 @ 2.8 GHz.  It is straightforward to further reduce the computation time by utilizing the conditional independency to parallelize the Gibbs sampling procedure.

Many downstream analyses can be conducted with this framework. In particular, URSM provides accurate estimates of the cell type specific profile matrix, which can be used for differential expression analysis between diseased and control samples. One can also apply URSM to single cells sequenced at different developmental periods to study the developmental trajectories of the cellular profiles.

As technologies improve and costs decline, single cell analysis can move to the new level by incorporating differential expression by maternal or paternal source of the chromosome. Such information can be captured if there are genetic differences between parents in the genes. Moreover genetic variation can affect expression of genes. Already experiments are being performed to determine which genetic variants are associated with changes in single cell expression.  This would allow analysis of expression based on parental origin of each copy of the gene.  These sources of variation are ignored in our model. Refining and extending scRNA-seq analytical tools to accommodate these sources of variation is one of the challenges for the future.

In this paper, we present our model assuming a given number of cell types $K$. In the situation where $K$ is not known a priori, one can first run the model using a larger value of $K$, examine the clustering of single cells after imputation, and then reduce to a reasonable choice of $K$ by combining cells from similar clusters.

Finally, we point out that the current model is developed under the setting of supervised learning where the labels for single cells are known. One can extend this framework to conduct  unsupervised cell clustering by introducing extra latent variables for cell labels in the hierarchical model. In addition, by the nature of the Multinomial distribution, the current model is fully determined by its first moment. Therefore, the imputation of single cell data may be further improved by introducing gene-gene correlations to the model. We leave the exploration in these directions to future work.

\section*{Acknowledgements}

We thank the anonymous reviewers and editor for their constructive comments and  suggestions. This work was supported by SF402281, SFARI124827, R37MH057881 (Kathryn Roeder and Bernie Devlin), and R01MH109900 (Kathryn Roeder), as well as DMS-1553884 and DMS-1407771 (Jing Lei).



\beginsupplement
\appendix

\section{Details of Gibbs sampling}
\label{appendix-gibbs}

Recall that the latent variables for bulk data and single cell data are conditionally independent given observed data $X, Y$ and parameters, so we discuss the Gibbs sampling algorithms for the two parts separately.

\subsection{Gibbs sampling for bulk data}

Recall that the bulk data part of the model is equivalent to the following mixture of multinomials:
\begin{align}
\begin{split}
&W_{\cdot j} \iid \textrm{Dirichlet}(\alpha)\,,~ j=1, \cdots, M\,, \\
&Z_{rj} \iid \textrm{Multinomial}(1, W_{\cdot j})\,, ~r=1, \cdots, R_j\,, \\
& d_{rj} \indep \textrm{Multinomial}(1, A_{\cdot Z_{rj}})\,, ~ r=1, \cdots, R_j\,, \\
& X_{ij} = \sum_{r=1}^{R_j} I_{\{d_{rj}=i\}}\,, ~ i=1, \cdots, N, ~ j=1, \cdots, M\,,
\end{split}
\end{align}
where $Z$ and $d$ are represented in scalars, i.e., $\Prob(Z_{rj}=k) = W_{kj}$, $\Prob(d_{rj}=i) = A_{i, Z_{rj}}$.
We further define
\begin{equation}
 \tilde{Z}_{ij, k} := \sum_{r: d_{rj}=i} I_{\{Z_{rj}=k\}}\,, ~~ \tilde{Z}_{ij} := (\tilde{Z}_{ij, k} ) \in \mathbb{R}^K\,,
 \end{equation}
then the complete likelihood function for bulk data can be written as below:
\begin{align}
\begin{split}
L_{bulk}( W, & Z,  d, X ~|~ \alpha, A)  \\
& = p(W|\alpha) p(Z|W) p(d|A, Z) p(X|d) \\
& = \prod_{j=1}^M \left\{  \Gamma \left(\sum_{t=1}^K \alpha_t\right) \prod_{k=1}^K \frac{W_{kj}^{(\alpha_k-1)}}{\Gamma(\alpha_k)} \cdot \prod_{r=1}^{R_j} \prod_{k=1}^K  W_{kj}^{I_{\{Z_{rj}=k\}}} \cdot  \right. \\
& \qquad \left. \cdot \prod_{k=1}^K \prod_{i=1}^N A_{ik}^{\sum_{r: d_{rj}=i} I_{\{Z_{rj}=k\}}} \cdot \prod_{i=1}^N I_{\{ X_{ij} = \sum_{r=1}^{R_j} I_{\{d_{rj}=i\}}  \}} \right\} \\
& =  \prod_{j=1}^M \left\{  \Gamma \left(\sum_{t=1}^K \alpha_t\right) \prod_{k=1}^K \frac{W_{kj}^{(\alpha_k-1)}}{\Gamma(\alpha_k)} \cdot \right. \\
& \qquad \left. \cdot \prod_{i=1}^{N} \prod_{k=1}^K \left( W_{kj} A_{ik} \right)^{\tilde{Z}_{ij, k} } \cdot  \prod_{i=1}^N I_{\{ X_{ij} = \sum_{r=1}^{R_j} I_{\{d_{rj}=i\}}  \}} \right\} \,.\\
\label{eq:ll-bulk}
\end{split}
\end{align}
Therefore,
\begin{align}
\begin{split}
& p(W_{\cdot j}~|~ \tilde{Z}, X, \alpha, A) \propto \prod_{k=1}^K W_{kj}^{ \left(\alpha_k + \sum_{i=1}^N \tilde{Z}_{ij, k} -1 \right)} \,, \\
& p(\tilde{Z}_{ij} ~|~ W, X, \alpha, A) \propto \left[ \prod_{k=1}^K \left( W_{kj} A_{ik} \right)^{\tilde{Z}_{ij, k}}\right] I_{\{ \sum_{k=1}^K \tilde{Z}_{ij, k} = X_{ij} \}}  \,.
\end{split}
\end{align}
and we immediately have
\begin{align}
\begin{split}
& W_{\cdot j} ~ | ~ \tilde{Z}, X \sim \textrm{Dirichlet} \left( \alpha + \sum_{i=1}^N \tilde{Z}_{ij} \right)\,, \\
& \tilde{Z}_{ij} ~ | ~ W, X \sim \textrm{Multinomial} \left( X_{ij}, \frac{ A_{i \cdot} \odot W_{\cdot j} }{ \sum_{k=1}^K A_{ik} W_{kj}}  \right)\,,
\end{split}
\end{align}
where $\odot$ denotes element-wise multiplication.

\subsection{Gibbs sampling for single cell data}
Recall that the first part of the single cell model can be rewritten as a  Bayesian Logistic regression as follows:
\begin{align}
\begin{split}
 & (\kappa_l, \tau_l) \sim N( \mu, \Sigma ) ,
 ~\textrm{where } \mu=(\mu_{\kappa}, \mu_{\tau}), \Sigma=\textrm{Diag}(\sigma_{\kappa}^2, \sigma_{\tau}^2)\,, \\
& S_{il}  \,|\,  \kappa_l, \tau_l \sim \textrm{Bernoulli}(\textrm{logistic}\left( \psi_{il}  \right))\,,~\textrm{where } \psi_{il} = \kappa_l + \tau_l A_{i, G_l}\,.
 \label{eq:bayes-logit}
 \end{split}
\end{align}
Therefore, we can utilize the data augmentation trick following  \cite{polson2013bayesian}. The key is to notice that the logistic function can be written as mixtures of Gaussians with respect to a Polya-Gamma (PG) distribution:
\begin{equation}
\frac{ (e^{\psi})^a }{ (1 + e^{\psi})^b }  = 2^{-b} e^{c \psi}  \int_0^{\infty} e^{- \omega \psi^2/2} p(\omega) d\omega, ~~ \forall \psi \in \mathbb{R} \,,
\end{equation}
for any constants $a, b>0$, where $c = a - b/2$ and $\omega \sim PG(0, 1)$. Plugging in this equation, and let
\[ \psi_{il} = \kappa_l + \tau_l A_{i, G_l}\,, \] 
the complete likelihood for the single cell data can be written as
\begin{align}
L_{sc}( & Y, \kappa, \tau, S ~|~ \mu_\kappa, \sigma_\kappa^2, \mu_\tau, \sigma_\tau^2, A) \nonumber \\
& = p(Y | S, A) p(S | \kappa, \tau) p(\kappa| \mu_\kappa, \sigma_\kappa^2) p(\tau|\mu_\tau, \sigma_\tau^2) \nonumber \\
& \propto \prod_{l=1}^L \left\{ \frac{R_l!}{\prod_{i=1}^N Y_{il}!} \prod_{i=1}^N \left[ \left( \frac{A_{i, G_l}}{ \sum_{n=1}^N S_{nl} A_{n, G_l} }  \right)^{Y_{il} S_{il}}  \delta_0(Y_{il})^{(1 - S_{il})} \right] \cdot \right.  \nonumber  \\
& \qquad \cdot \left. \left[ \prod_{i=1}^N \frac{ \left( e^{\psi_{il}} \right)^{S_{il}} }{ 1 + e^{\psi_{il}}} \right] \cdot \frac{1}{\sigma_\kappa  \sigma_\tau} \exp\left\{  - \frac{ (\kappa_l - \mu_\kappa)^2 }{2 \sigma_\kappa^2 } - 
\frac{ (\tau_l - \mu_\tau)^2 }{ \sigma_\tau^2 }  \right\} \right\}  \nonumber\\
& \propto \prod_{l=1}^L \left\{  \left[ \frac{R_l!}{\prod_{i=1}^N Y_{il}!} \frac{ \prod_{i=1}^N \left(A_{i, G_l}  \right)^{Y_{il} S_{il}}  \delta_0(Y_{il})^{(1 - S_{il})} }{ \left(  \sum_{n=1}^N S_{nl} A_{n, G_l} \right)^{R_l} } \right] \cdot \right. \nonumber \\
& \qquad \cdot \left. \left[ \prod_{i=1}^N e^{\psi_{il} (S_{il} - 1/2)}  \int_0^{\infty} e^{- \omega_{il} \psi_{il}^2/2} p(\omega_{il}) d\omega_{il} \right] \cdot \right. \nonumber \\
& \qquad \cdot \left. (\sigma_\kappa ^2  \sigma_\tau^2)^{- 1/2} \exp\left\{  - \frac{ (\kappa_l - \mu_\kappa)^2 }{2 \sigma_\kappa^2 } - 
\frac{ (\tau_l - \mu_\tau)^2 }{ 2\sigma_\tau^2 }  \right\} \right\} \nonumber \\ 
& \propto \int_0^{\infty} p(Y, \kappa, \tau, S, \omega) p(\omega) d\omega
 \label{eq:ll-sc}
\end{align}
where $\delta_0(y) = I_{\{y=0\}}$, and $\omega_{il} \sim PG(0, 1)$ independently.
Then following the same arguments as in \cite{polson2013bayesian}, we get the conditional posterior distribution for $\omega_{il}, \kappa_l, \tau_l$ as follows:
\begin{align}
\begin{split}
& \omega_{il} \,|\, \omega_{-(il)}, S, Y, \kappa, \tau ~ \sim ~ \textrm{Polya-Gamma}(1, \psi_{il}) \,, \\
& (\kappa_l, \tau_l) \,|\, \omega, S, Y  ~ \sim ~  N(m_{\omega l}, V_{\omega l}^{-1}) \,, \\
\end{split}
\end{align}
where
\begin{gather*}
V_{\omega l} = \left(\begin{array}{cc}
\sum_{i=1}^N \omega_{il} + \sigma_{\kappa}^{-2} & \sum_{i=1}^N \omega_{il} A_{i, G_l} \\
\sum_{i=1}^N \omega_{il} A_{i, G_l} & \sum_{i=1}^N \omega_{il} A_{i, G_l}^2 + \sigma_{\tau}^{-2}
\end{array}\right)
, \\
 m_{\omega l} = V_{\omega l}^{-1} 
 \left(\begin{array}{c}
 \sum_{i=1}^N S_{il} - N/2 + \mu_\kappa / \sigma_\kappa^2 \\
 \sum_{i=1}^N S_{il} A_{i, G_l} - 1/2 + \mu_\tau / \sigma_\tau^2 
 \end{array}\right)\,.
\end{gather*}
The only thing left is the conditional posterior for $S_{il}$. This can be easily obtained by looking at the un-augmented version of likelihood. Note that $S_{il}$ is binary, and we have
\begin{equation}
 \frac{P(S_{il}=1 ~ | ~ S_{-(il)}, Y, \psi)} {  P(S_{il}=0 ~ | ~ S_{-(il)}, Y, \psi) } = \frac{ (A_{i, G_l})^{Y_{il}} e^{\psi_{il}} }{ \delta_0(Y_{il})  } \cdot \left(\frac{ \sum_{n\neq i}^N S_{nl} A_{n, G_l} }{ \sum_{n\neq i}^N S_{nl} A_{n, G_l} + A_{i, G_l}}\right)^{R_l} \,. 
 \end{equation} 
Therefore,
\[ S_{il} \,|\, S_{-(il)}, \omega, \kappa, \tau, Y ~ \sim ~ \textrm{Bernoulli}(b_{il})\,,
 \]
 where
 \[
 b_{il} = \begin{cases}
1, & \textrm{if } Y_{il} > 0 \\
\textrm{logit}\left( \psi_{il} + R_l \log \left( \frac{ \sum_{n \neq i} S_{nl}A_{n, G_l}} { A_{i, G_l}  + \sum_{n \neq i} S_{nl} A_{n, G_l} } \right) \right), & \textrm{if } Y_{il}=0
\end{cases}
\,.
 \]

\section{Details of the EM algorithm}
\label{appendix-mstep}

\subsection{M-step in EM algorithm}
Here we give more details about the M-step in the Gibbs-EM (GEM) algorithm. By combining equations \cref{eq:ll-bulk} and \cref{eq:ll-sc}, the expectation of  complete log likelihood function can be easily derived. However, the term of $\Exp \left[ \log\left(\sum_{n=1}^N S_{nl} A_{n, G_l}\right) \right]$ makes the optimization complicated. We work around this issue by optimizing a {\it lower bound} of the objective function. The key step is the following lower bound using Jensen's inequality \citep{paisley2013two-useful}:
\begin{equation}
 - \Exp \left[ \log \left( \sum_n X_n  \right) \right]  \geq - \log u -  \frac{\sum_{n} \Exp [X_n] - u}{u} \,,
 \label{eq:jensen}
 \end{equation}
where $u = \sum_{n} \Exp[X_n]$, for any random variables $X_n$'s. Using inequality \cref{eq:jensen}, we get the following lower bound of the expected complete log likelihood, using the augmented version:

\begin{align}
\begin{split}
\Exp_Q & \left[ \log p(X, Y, W, \tilde{Z}, \kappa, \tau, S, \omega | \theta) \right] \geq \textrm{const.} + \\
&  \sum_{j=1}^M \left\{ \log \Gamma \left(\sum_{k=1}^K \alpha_k \right) + \sum_{k=1}^K  \left[ (\alpha_k-1) \Exp_Q\left[\log W_{kj}\right]  - \log \Gamma(\alpha_k) \right]   \right. \\
& \quad \left. +\sum_{i=1}^N \sum_{k=1}^K \left[ \Exp_Q \left( \tilde{Z}_{ij, k} \log W_{kj} \right) + \Exp_Q \left( \tilde{Z}_{ij, k} \right) \log A_{ik} \right] \right\} + \\
& \sum_{l=1}^L \left\{ \sum_{i=1}^N \Exp_Q(S_{il})  Y_{il} \log (A_{i, G_l}) - 
R_l \left( \sum_{i=1}^N \frac{\Exp_Q(S_{il}) A_{i, G_l}}{u_l} + \log u_l \right)  \right. \\
 &\quad  \left.  +\sum_{i=1}^N \Exp_Q\left[ \left(S_{il} - \frac{1}{2} \right)(\kappa_l + \tau_l A_{i, G_l})  - \frac{ \omega_{il} (\kappa_l + \tau_l A_{i, G_l})^2 }{2} \right] \right. \\
 & \quad \left. - \frac{1}{2} \left( \log \sigma_\kappa^2 + \log \sigma_\tau^2 \right)  - \frac{ \Exp_Q \left[ (\kappa_l - \mu_\kappa)^2 \right] }{2 \sigma_\kappa^2 } - 
\frac{ \Exp_Q \left[ (\tau_l - \mu_\tau)^2 \right] }{ 2\sigma_\tau^2 }  \right\}  \,,
\label{eq:ll-full}
\end{split}
\end{align}
where $u_l = \sum_{i=1}^N A_{i, G_l} \Exp_Q(S_{il})$, and $\Exp_Q$ is the expectation of the posterior distribution, which can be estimated using Gibbs samples.  We omit the constants that only involve the data $(X, Y)$, since the goal here is to optimize over the parameters $\theta=(A, \alpha, \mu_\kappa, \sigma_\kappa^2, \mu_\tau, \sigma_\tau^2)$. In addition, we use the fact that 
\begin{equation}
 \Exp_Q[S_{il}] = 1 ~~ \textrm{when} ~~ Y_{il} > 0\,. 
 \end{equation}
The above lower bound is also referred to as the Evidence Lower BOund (ELBO). With  \cref{eq:ll-full}, it is straightforward to derive the derivatives mentioned in equation (3.9) in the original manuscript.

The final missing piece is the projection functions to the feasible set. The projection function for $\alpha$ is straightforward: for any constant $\epsilon_{\alpha} >0$,
\begin{equation}
\textrm{Proj} (\alpha_k) = \max\{\epsilon_{\alpha}, ~ \alpha_k\}\,.
\end{equation}
As for the profile matrix $A$, the projection function is to project onto a subset of simplex
\begin{equation}
 \mathcal{S}_\epsilon = \left\{ u= (u_1, \cdots, u_N) \in \mathbb{R}^N: \sum_{n=1}^N u_n = 1, ~ u_n \geq \epsilon, ~ \forall n  \right\}\,, 
 \end{equation}
for some constant $\epsilon > 0$. \cite{wang2013projection}  propose an efficient algorithm for the case when $\epsilon = 0$. The algorithm can be easily generalized to handle a general $\epsilon$, which is specified below:
\paragraph{Projection algorithm for $A$} For any vector $v \in \mathbb{R}^N$, constant $\epsilon \geq 0$,
\begin{enumerate}
\item Sort $v$ into $\tilde{v}$, such that $\tilde{v}_1\geq \tilde{v}_2 \geq \cdots \geq \tilde{v}_N$.
\item Find $\rho = \max \left\{1 \leq j \leq N: \tilde{v}_j + \frac{1}{j}( 1 - \sum_{i=1}^j \tilde{v}_i - (N-j) \epsilon )  > \epsilon \right\}$.
\item Let $\lambda = \frac{1}{\rho} \left( 1 - \sum_{i=1}^\rho \tilde{v}_i - (N-\rho) \epsilon \right)$.
\item Let $v^*_i = \max \{ \tilde{v}_i + \lambda, ~ \epsilon \}$, then $v^* = (v^*_i) \in \mathbb{R}^N$ is the projection.
\end{enumerate}

\subsection{Starting values}

Here we present some heuristic for choosing the starting values for the EM algorithm.
For  the profile matrix $A$, a good candidate is the sample means $\hat{A}^{naive}$ in single cell data:
\begin{equation} 
\hat{A}_{ik}^{naive} = \frac{1}{ \# \{l: G_l=k\}} \sum_{l: G_l=k} \frac{Y_{il}}{R_l} \,.
\label{eq:samplemean}
\end{equation}
As illustrated in Figure 2b of  the original manuscript, although being biased, $\hat{A}^{naive}$ is usually not too far away from the true profile matrix $A$. 

For $\alpha$, the starting value can be chosen using prior knowledge. Although the exact mixing proportions for each bulk sample is unknown, scientists usually have a good sense of the rough proportions of different cell types in certain tissues. In the case where prior knowledge is unavailable, $\alpha$ can simply be set to $\mathbf{1}^K$, which corresponds to a uniform distribution. 
In fact, all the simulations in the original manuscript use $\alpha_0=(1,1,1)$ as the starting value, and the performances are satisfactory. 

Finally, the starting values of $\mu_\kappa$ and $\mu_\tau$ can be set according to $\hat{A}_{ik}^{naive}$, such that the distribution of $\left\{ 1 - \textrm{logistic} \left(\kappa_l + \tau_l \hat{A}_{i, G_l}^{naive} \right) \right\}_{i, l}$ matches our prior knowledge of the dropout probabilities. In our simulation studies, $\mu_\kappa$ is initialized at $\mu_{\kappa, 0} = \textrm{logit}(0.4)$,  so that the maximal dropout probability is $60\%$ (achieved when $A_{il} = 0$), and $\mu_\tau$ is initialized at
$
\mu_{\tau, 0} =  (\textrm{logit}(0.7) - \mu_{\kappa, 0}) / \bar{A}^{naive}
$
where $\bar{A}^{naive} = \sum_{i, k} \hat{A}_{ik}^{naive} / (NK)$. Therefore, using the initial values, the empirical average dropout probabilities is $ \frac{1}{NK} \sum_{i, k} ( \mu_{\kappa, 0} + \mu_{\tau,0} \hat{A}_{i, k}^{naive} )  = 30\%$.

\subsection{A MAP approximation} 
The bottleneck of the computation is the Gibbs sampling step in the EM algorithm.
Here, we present a fast algorithm that avoids the Gibbs sampling for bulk samples by plugging in the maximum a posteriori (MAP) estimation of $W$. This algorithm is useful when dealing with larger data sets where computation load becomes a major concern. The simulation results in Section 4.5 of the original manuscript are obtained using this MAP approximation.

Specifically, we further assume that the sequencing depth of bulk samples is generated from $R_j \sim \textrm{Poisson}(\lambda)$. Then the distribution of $X$ after marginalizing out $R_j$ becomes
\[ X_{ij} | A,   W \sim \textrm{Poisson}(\lambda (AW)_{ij})\,, \]
which gives
\[ p(X, W | A, \alpha) \propto \prod_{i=1}^N \prod_{j=1}^M \frac{ \left((AW)_{ij} \lambda \right)^{X_{ij}} \exp \{- \lambda (AW)_{ij}\} }{ X_{ij}! }  \cdot \prod_{j=1}^M \prod_{k=1}^K W_{kj}^{(\alpha_k - 1)}\,.\]
Therefore,  the MAP of $W$ can be obtained by
\[\begin{split}
& \max_W \sum_{i, j} \left[ X_{ij} \log((AW)_{ij}) - \lambda (AW)_{ij}  \right] + \sum_{k, j} (\alpha_k - 1) \log W_{kj} \,, \\
& \textrm{subject to } \sum_{k} W_{kj} = 1 \,. 
 \end{split}\]
To get the optimal $W_{kj}$, we write down the KKT conditions:
\[\begin{cases}
 \sum_i \frac{X_{ij} A_{ik}}{(AW)_{ij}} - \lambda + \frac{\alpha_k - 1} {W_{kj}} + \eta_j = 0\,, \\
 \sum_{k} W_{kj} = 1\,, 
\end{cases}\]
where $\eta_j$ is the Lagrangian multiplier for the constraint $\sum_{k} W_{kj} = 1$.
Finally, we obtain the following fixed-point iteration for $W$:
\begin{equation}\begin{split}
W_{kj} \propto W_{kj}  \sum_i \frac{X_{ij} A_{ik}}{(AW)_{ij}} + \alpha_k - 1 \,, ~~ 
\textrm{s.t. }  \sum_{k} W_{kj} = 1 
\end{split}
\label{eq:w-update}
\end{equation}
In our implementation of this fast algorithm, we perform one update of equation \cref{eq:w-update} in each E-step for bulk samples, starting from the $W$ obtained from the previous EM iteration. Interestingly, if $\alpha_k = 1$, this algorithm recovers the multiplicative updates for non-negative matrix factorization when the divergence loss is used \citep{lee2001algorithms}.

\bibliography{singlecell}
\bibliographystyle{imsart-nameyear}

\end{document}